\input ./style/arxiv-general.cfg
\documentclass[MSNbibl,number,seceqn,dvips]{arxstspdf}
\makeatletter
   \@ifpackageloaded{graphicx}{}{\usepackage{graphicx}}
\makeatother
\usepackage{algorithm}
\usepackage{flushend}
\usepackage{stfloats}
\usepackage{url,breakurl}

\volume{30}
\issue{3}
\pubyear{2015}
\firstpage{328}
\lastpage{351}
\doi{10.1214/14-STS511}
\docsubty{FLA}

\makeatletter
\newcommand{\rrvert}{\vert}
\newcommand{\rrVert}{\Vert}
\newcommand{\llvert}{\vert}
\newcommand{\llVert}{\Vert}
\renewcommand{\mid}{|}
\makeatother

\begin{document}
\begin{frontmatter}

\title{On Particle Methods for Parameter Estimation in State-Space
Models}
\runtitle{On Particle Methods for Parameter Estimation in State-Space Models}

\begin{aug}
\author[A]{\fnms{Nikolas}~\snm{Kantas}\ead[label=e1]{n.kantas@imperial.ac.uk}},
\author[B]{\fnms{Arnaud}~\snm{Doucet}\corref{}\ead
[label=e2]{doucet@stats.ox.ac.uk}},
\author[C]{\fnms{Sumeetpal S.}~\snm{Singh}\ead[label=e3]{sss40@eng.cam.ac.uk}},
\author[C]{\fnms{Jan}~\snm{Maciejowski}\ead[label=e4]{jmm@eng.cam.ac.uk}}
\and
\author[D]{\fnms{Nicolas}~\snm{Chopin}\ead[label=e5]{nicolas.chopin@ensae.fr}}
\runauthor{N. Kantas et al.}
%
\affiliation{Imperial College London, University of Oxford, University
of Cambridge, University of Cambridge and CREST-ENSAE \& HEC}
\address[A]{N.~Kantas is Lecturer, Department of Mathematics, Imperial
College London, London SW7 2BZ, United Kingdom \printead{e1}.}
\address[B]{A.~Doucet is Professor, Department of Statistics, University
of Oxford, Oxford OX1 3TG, United Kingdom \printead{e2}.}
\address[C]{\mbox{S.~S.~Singh} is Senior Lecturer, Department of Engineering, University of Cambridge,
Cambridge CB1 2PZ, United Kingdom \printead{e3}.
\mbox{J.~M.~Maciejowski} is Professor, Department of Engineering, University of
Cambridge, Cambridge CB1 2PZ, United Kingdom  \printead{e4}.}
\address[D]{N.~Chopinis Professor, CREST-ENSAE and HEC Paris, 99245 Malakoff,
France \printead{e5}.}
\end{aug}

%
\begin{abstract}
Nonlinear non-Gaussian state-space models are ubiquitous in statistics,
econometrics, information engineering and signal processing. Particle
methods, also known as Sequential Monte Carlo (SMC) methods, \mbox{provide}
reliable numerical approximations to the associated state inference
problems. However, in most applications, the state-space model of
interest also depends on unknown static parameters that need to be
estimated from the data. In this context, standard particle methods
fail and it is necessary to rely on more sophisticated algorithms.
The aim of this paper is to present a comprehensive review of particle
methods that have been proposed to perform static parameter estimation
in state-space models. We discuss the advantages and limitations of
these methods and illustrate their performance on simple models.
\end{abstract}

%
\end{frontmatter}

\section{Introduction}

State-space models, also known as hidden Markov models, are a very
popular class of time series models that have found numerous of applications
in fields as diverse as statistics, ecology, econometrics, engineering
and environmental sciences; see \cite
{cappe2005,Douc01,Elliott1996,west97}. Formally, a state-space model
is defined by two stochastic processes $\{X_{n}\}_{n\geq0}$ and $\{
Y_{n}\}_{n\geq0}$.
The process $\{X_{n}\}_{n\geq0}$ is an $\mathcal{X}$-valued latent
Markov process of initial density $\mu_{\theta} (x )$ and
Markov transition density\ $f_{\theta}(x^{\prime}\mid x)$, that is,
%
\begin{eqnarray}\label{EqustateSpaceSystemLitRev1}
&& X_{0}\sim\mu_{\theta} (x_{0} ),
\nonumber\\[-8pt]\\[-8pt]\nonumber
&&X_{n}\llvert(X_{0\dvtx n-1}=x_{0\dvtx n-1} )\sim f_{\theta}
(x_{n}\rrvert x_{n-1}),
\end{eqnarray}
whereas the $\mathcal{Y}$-valued observations $\{Y_{n}\}_{n\geq0}$
satisfy
%
\begin{eqnarray}\label{EqustateSpaceSystemLitRev-2}
&& Y_{n}|(X_{0\dvtx n}=x_{0\dvtx n},
Y_{0\dvtx n-1}=y_{0\dvtx n-1} )
\nonumber\\[-8pt]\\[-8pt]\nonumber
&&\quad \sim g_{\theta}(y_{n}\mid x_{n}),
\end{eqnarray}
where $g_{\theta}(y\mid x)$ denotes the conditional marginal density,
$\theta\in\Theta$ the parameter of the model and $z_{i\dvtx j}$ denotes
components $ (z_{i},z_{i+1},\ldots,z_{j} )$ of a sequence
$ \{ z_{n} \} $.
The spaces $\mathcal{X}$ and $\mathcal{Y}$ can be Euclidean, but
what follows applies to more general state spaces as well.

The popularity of state-space models stems from the fact that they
are flexible and easily interpretable. Applications of state-space
models include stochastic volatility models where $X_{n}$ is the
volatility of an asset and $Y_{n}$ its observed log-return \cite
{KimShephardChib},
biochemical network models where $X_{n}$ corresponds to the population
of various biochemical species and $Y_{n}$ are imprecise measurements
of the size of a subset of these species \cite{wilkinson2012}, neuroscience
models where $X_{n}$ is a state vector determining the neuron's
stimulus--response
function and $Y_{n}$ some spike train data \cite{paninski2010}.
However, nonlinear non-Gaussian state-space models are also notoriously
difficult to fit to data and it is only recently, thanks to the advent
of powerful simulation techniques, that it has been possible to fully
realize their potential.

To illustrate the complexity of inference in state-space models, consider
first the scenario where the parameter $\theta$ is \emph{known}.
On-line and off-line inference about the state process $\{X_{n}\}$
given the observations $\{Y_{n}\}$ is only feasible analytically
for simple models such as the linear Gaussian state-space model. In
nonlinear non-Gaussian scenarios, numerous approximation schemes, such
as the Extended Kalman filter or the Gaussian sum filter \cite{gaussiansum},
have been proposed over the past fifty years to solve these so-called
optimal filtering and smoothing problems, but these methods lack rigor
and can be unreliable in practice in terms of accuracy, while deterministic
integration methods are difficult to implement. Markov chain Monte
Carlo (MCMC) methods can obviously be used, but they are impractical
for on-line inference; and even for off-line inference, it can be difficult
to build efficient high-dimensional proposal distributions for such
algorithms. For nonlinear non-Gaussian state-space models \emph{particle
algorithms} have emerged as the most successful. Their widespread
popularity is due to the fact that they are easy to implement, suitable
for parallel implementation \cite{LeeWhiteley2014forest} and, more
importantly, have been demonstrated in numerous settings to yield
more accurate estimates than the standard alternatives, for example,
see \cite{cappe2005,delmoral2004,Douc01,Liu01}.

In most practical situations, the model (\ref
{EqustateSpaceSystemLitRev1})--(\ref{EqustateSpaceSystemLitRev-2})
depends on an \emph{unknown} parameter vector $\theta$ that needs
to be inferred from the data either in an on-line or off-line manner.
In fact inferring the parameter $\theta$ is often the primary problem
of interest; for example, for biochemical networks, we are not interested
in the population of the species per se, but we want to infer some
chemical rate constants, which are parameters of the transition prior
$f_{\theta}(x^{\prime}\mid x)$. Although it is possible to define an
extended state that includes the original state $X_{n}$ and the parameter
$\theta$ and then apply standard particle methods to perform parameter
inference, it was recognized very early on that this naive approach
is problematic \cite{Kita98} due to the parameter space not being
explored adequately. This has motivated over the past fifteen years
the development of many particle methods for the parameter estimation
problem, but numerically robust methods have only been proposed recently.
The main objective of this paper is to provide a comprehensive overview
of this literature. This paper thus differs from recent survey papers
on particle methods which all primarily focus on estimating the state
sequence $X_{0\dvtx n}$ or discuss a much wider range of topics, for example,
\cite{doucet2011,kitagawa2014,kunsch2013,LinLiuChen2012}.
We will present the main features of each method and comment on their
pros and cons. No attempt, however, is made to discuss the intricacies
of the specific implementations. For this we refer the reader to the
original references.

We have chosen to broadly classify the methods as follows: Bayesian
or Maximum Likelihood (ML) and whether they are implemented off-line
or on-line. In the Bayesian approach, the unknown parameter is assigned
a prior distribution and the posterior density of this parameter given
the observations is to be characterized. In the ML approach, the
parameter estimate is the maximizing argument of the likelihood of
$\theta$ given the data. Both these inference procedures can be carried
out off-line or on-line. Specifically, in an off-line framework we
infer $\theta$ using a fixed observation record $y_{0\dvtx T}$. In contrast,
on-line methods update the parameter estimate sequentially as observations
$ \{ y_{n} \} _{n\geq0}$ become available.

The rest of the paper is organized as follows. In Section~\ref
{seccomputationalissues}
we present the main computational challenges associated to parameter
inference in state-space models. In Section~\ref{secfilteringandparticle}
we review particle methods for filtering when the model does not
include any unknown parameters, whereas Section~\ref{secsmoothingandparticle}
is dedicated to smoothing. These filtering and smoothing techniques
are at the core of the off-line and on-line ML parameter procedures
described in Section~\ref{secMLestimation}. In Section~\ref
{secBayesianestimation}
we discuss particle methods for off-line and on-line Bayesian parameter
inference. The performance of some of these algorithms are illustrated
on simple examples in Section~\ref{secexperimentalresults}. Finally,
we summarize the main advantages and drawbacks of the methods presented
and discuss some open problems in Section~\ref{secconclusion}.

\section{Computational Challenges Associated to Parameter Inference}
\label{seccomputationalissues}

A key ingredient of ML and Bayesian parameter inference is the likelihood
function $p_{\theta} (y_{0\dvtx n} )$ of $\theta$ which satisfies
%
\begin{equation}
p_{\theta} (y_{0\dvtx n} )=\int p_{\theta} (x_{0\dvtx n},y_{0\dvtx n}
)\,dx_{0\dvtx n},\label{eqdecompositionjointdistribution}
\end{equation}
where $p_{\theta} (x_{0\dvtx n},y_{0\dvtx n} )$ denotes the joint
density of $ (X_{0\dvtx n},Y_{0\dvtx n} )$ which is given from equations
(\ref{EqustateSpaceSystemLitRev1})--(\ref{EqustateSpaceSystemLitRev-2})
by
%
\begin{eqnarray}\label{eqjointpdfxy}
&& p_{\theta} (x_{0\dvtx n},y_{0\dvtx n} )
\nonumber\\[-8pt]\\[-10pt]\nonumber
&&\quad =\mu_{\theta}
(x_{0} )\prod_{k=1}^{n}f_{\theta}
(x_{k}\mid x_{k-1} )\prod
_{k=0}^{n}g_{\theta}
(y_{k}\mid x_{k} ).\hspace*{-10pt}
\end{eqnarray}
The likelihood function is also the normalizing constant of the posterior
density $p_{\theta} (x_{0\dvtx n}\mid y_{0\dvtx n} )$ of the latent states
$X_{0\dvtx n}$ given data $y_{0\dvtx n}$,
%
\begin{equation}
p_{\theta} (x_{0\dvtx n}\mid y_{0\dvtx n} )=\frac{p_{\theta}
(x_{0\dvtx n},y_{0\dvtx n} )}{p_{\theta} (y_{0\dvtx n} )}.\label
{eqfilteringdensityjoint}
\end{equation}
This posterior density is itself useful for computing the score vector
$\nabla_{\theta}\ell_{n} (\theta)$ associated to the
log-likelihood
$\ell_{n}(\theta)=\log p_{\theta}
(y_{0\dvtx n} )$,
as Fisher's identity yields
%
\begin{eqnarray}\label{eqfisheridentityloglike}
\nabla_{\theta}\ell_{n} (\theta) &=& \int\nabla_{\theta
}\log p_{\theta} (x_{0\dvtx n},y_{0\dvtx n} )
\nonumber\\[-8pt]\\[-8pt]\nonumber
&&\hspace*{9pt}{}\cdot p_{\theta}
(x_{0\dvtx n}\mid y_{0\dvtx n} )\,dx_{0\dvtx n}.
\end{eqnarray}

The main practical issue associated to parameter inference in nonlinear
non-Gaussian state-space models is that the likelihood function is
intractable. As performing ML parameter inference requires maximizing
this intractable function, it means practically that it is necessary
to obtain reasonably low-variance Monte Carlo estimates of it, or
of the associated score vector if this maximization is carried out
using gradient-based methods. Both tasks involve approximating
high-dimensional integrals, (\ref{eqdecompositionjointdistribution})
and (\ref{eqfisheridentityloglike}), whenever $n$ is large. On-line
inference requires additionally that these integrals be approximated
on the fly, ruling out the applications of standard computational
tools such as MCMC.

Bayesian parameter inference is even more challenging, as it requires
approximating the posterior density
%
\begin{eqnarray}
p(\theta\mid y_{0\dvtx n}) & = & \frac{p_{\theta}(y_{0\dvtx n})p(\theta)}{\int
p_{\theta}(y_{0\dvtx n})p(\theta)\,d\theta},\label{eqParameterPosterior}
\end{eqnarray}
where $p(\theta)$ is the prior density. Here not only $p_{\theta}(y_{0\dvtx n})$
but also $p (y_{0\dvtx n} )=\int p_{\theta}(y_{0\dvtx n})p(\theta
)\,d\theta$
are intractable and, once more, these integrals must be approximated
on-line if one wants to update the posterior density sequentially.
We will show in this review that particle methods are particularly
well suited to these integration tasks.

\section{Filtering and Particle Approximations}\label{secfilteringandparticle}

In this section the parameter $\theta$ is assumed known and we focus
on the problem of estimating the latent process $\{X_{n}\}_{n\geq0}$
sequentially given the observations. An important by-product of this
so-called filtering task from a parameter estimation viewpoint is
that it provides us with an on-line scheme to compute $ \{
p_{\theta} (y_{0\dvtx n} ) \} _{n\geq0}$.
As outlined in Section~\ref{seccomputationalissues}, the particle
approximation of these likelihood terms is a key ingredient of numerous
particle-based parameter inference techniques discussed further on.

\subsection{Filtering}\label{secfiltering}

Filtering usually denotes the task of estimating recursively in time
the sequence of marginal posteriors $ \{ p_{\theta} (x_{n}\mid y_{0\dvtx n}
) \} _{n\geq0}$,
known as the filtering densities. However, we will adopt here a more
general definition and will refer to filtering as the task of estimating
the sequence of joint posteriors $ \{ p_{\theta} (x_{0\dvtx n}\mid y_{0\dvtx n} )
\} _{n\geq0}$
recursively in time, but we will still refer to the marginals $ \{
p_{\theta} (x_{n}\mid y_{0\dvtx n} ) \}
_{n\geq0}$
as the filtering densities.

\begin{algorithm*}
\begin{itemize}
\item\emph{At time} $n=0$, for all $i\in\{ 1,\ldots,N\} $:
\begin{enumerate}
\item\textsf{Sample }$X_{0}^{i}\sim q_{\theta}( x_{0}\mid y_{0})$.
\item\textsf{Compute }$\overline{W}_{1}^{i}\propto w_{0}
(X_{0}^{i} )q_{\theta} (y_{1}\mid X_{0}^{i} )$,
$\sum_{i=1}^{N}\overline{W}_{1}^{i}=1$.
\item\textsf{Resample }$\overline{X}_{0}^{i}\sim\sum_{i=1}^{N}\overline
{W}_{1}^{i}\delta_{X_{0}^{i}} (dx_{0} )$.
\end{enumerate}
\item\emph{At time} $n\geq1$, for all $i\in\{ 1,\ldots,N \} $:
\begin{enumerate}
\item\textsf{Sample }$X_{n}^{i}\sim q_{\theta}( x_{n}\mid
y_{n},\overline{X}_{n-1}^{i})$
\textsf{and set} $X_{0\dvtx n}^{i}\leftarrow(\overline
{X}_{0\dvtx n-1}^{i},X_{n}^{i} )$.
\item\textsf{Compute }$\overline{W}_{n+1}^{i}\propto w_{n}
(X_{n-1\dvtx n}^{i} )q_{\theta} (y_{n+1}\mid X_{n}^{i} )$,
$\sum_{i=1}^{N}\overline{W}_{n+1}^{i}=1$.
\item\textsf{Resample }$\overline{X}_{0\dvtx n}^{i}\sim\sum
_{i=1}^{N}\overline{W}_{n+1}^{i}\delta_{X_{0\dvtx n}^{i}}
(dx_{0\dvtx n} )$\textsf{.}
\end{enumerate}
\end{itemize}
\protect\caption{\textbf{Auxiliary particle filtering}}
\label{algAPF}
\end{algorithm*}

It is easy to verify from (\ref{eqdecompositionjointdistribution})
and (\ref{eqfilteringdensityjoint}) that the posterior $p_{\theta
} (x_{0\dvtx n}\mid y_{0\dvtx n} )$
and the likelihood $p_{\theta} (y_{0\dvtx n} )$ satisfy the following
fundamental recursions: for $n\geq1$,
%
\begin{eqnarray}\label{eqjointupdate}
&& p_{\theta} (x_{0\dvtx n}\mid y_{0\dvtx n} )
\nonumber\\[-8pt]\\[-8pt]\nonumber
&&\quad =p_{\theta
}(x_{0\dvtx n-1}\mid y_{0\dvtx n-1} )
\frac{f_{\theta
} (x_{n}\mid x_{n-1} )g_{\theta} (y_{n}\mid x_{n} )}{p_{\theta}
(y_{n}\mid
y_{0\dvtx n-1} )}\hspace*{-20pt}
\end{eqnarray}
and
%
\begin{equation}
p_{\theta} (y_{0\dvtx n} )=p_{\theta} (y_{0\dvtx n-1}
)p_{\theta} (y_{n}\mid y_{0\dvtx n-1} ),\label
{eqlikelihoodDecomp}
\end{equation}
where
%
\begin{eqnarray}\label{eqrecursivelikeliDecomp}
&& p_{\theta} ( y_{n}\mid y_{0\dvtx n-1} )
\nonumber
\\
&&\quad =\int g_{\theta}
( y_{n}\mid x_{n} )f_{\theta
} ( x_{n}\mid
x_{n-1} )
\\
&&\hspace*{30pt}{} \cdot p_{\theta} (x_{n-1}\mid y_{0\dvtx n-1}
)\,dx_{n-1\dvtx n}.\nonumber
\end{eqnarray}

There are essentially two classes of models for which $p_{\theta}
(x_{0\dvtx n}\mid y_{0\dvtx n} )$
and $p_{\theta} (y_{0\dvtx n} )$ can be computed exactly:  the
class of linear Gaussian models, for which the above recursions may
be implemented using Kalman techniques, and when $\mathcal{X}$ is
a finite state space; see, for example, \cite{cappe2005}. For other
models these quantities are typically intractable, that is, the densities
in (\ref{eqjointupdate})--(\ref{eqrecursivelikeliDecomp}) cannot
be computed exactly.

\subsection{Particle Filtering}\label{secsmc-filtering}

\subsubsection{Algorithm}

Particle filtering methods are a set of simulation-based techniques
which approximate numerically the recursions (\ref{eqjointupdate})
to (\ref{eqrecursivelikeliDecomp}). We focus here on the APF (auxiliary
particle filter \cite{Pitt99}) for two reasons:  first, this is a
popular approach, in particular, in the context of parameter estimation
(see, e.g., Section~\ref{subUsing-MCMCsteps-within-SMC}); second,
the APF covers as special cases a large class of particle algorithms,
such as the bootstrap filter \cite{Itallstartedwith} and SISR
(Sequential Importance Sampling Resampling \cite{Douc00,Liu98}).

Let
%
\begin{equation}\label{eqAPFproposal}
q_{\theta} (x_{n},y_{n}\mid x_{n-1}
)=q_{\theta} (x_{n}\mid y_{n},x_{n-1}
)q_{\theta} (y_{n}\mid x_{n-1} ),\hspace*{-30pt}
\end{equation}
where $q_{\theta} (x_{n}\mid y_{n},x_{n-1} )$ is a probability
density function which is easy to sample from and $q_{\theta}
(y_{n}\mid x_{n-1} )$
is not necessarily required to be a probability density function but
just a nonnegative function of $ (x_{n-1},y_{n} )\in
\mathcal{X}\times\mathcal{Y}$
one can evaluate. [For $n=0$, remove the dependency on $x_{n-1}$,
i.e., $q_{\theta}(x_{0},y_{0})=q_{\theta}(x_{0}\mid y_{0})q_{\theta}(y_{0})$.]

The algorithm relies on the following importance weights:
%
\begin{eqnarray}
w_{0} (x_{0} ) & =&\frac{g_{\theta} (y_{0}\mid x_{0}
)\mu_{\theta} (x_{0} )}{q_{\theta} (x_{0}\mid y_{0} )},\label
{eqimportanceweightinit}
\\
w_{n} (x_{n-1\dvtx n} ) & =& \frac{g_{\theta}
(y_{n}\mid x_{n} )f_{\theta} (x_{n}\mid x_{n-1} )}{q_{\theta
} (x_{n},y_{n}\mid x_{n-1} )}
\nonumber\\[-8pt]\label{eqimportanceweights} \\[-8pt]
\eqntext{\mbox{for }n \geq1.}
\end{eqnarray}
In order to alleviate the notational burden, we omit the dependence
of the importance weights on $\theta$; we will do so in the remainder
of the paper when no confusion is possible. The auxiliary particle
filter can be summarized in Algorithm \ref{algAPF} \cite{fearnhead1999,Pitt99}.


One recovers the SISR algorithm as a special case of Algorithm \ref{algAPF}
by taking $q_{\theta}(y_{n}\mid x_{n-1})=1$ [or, more generally, by taking
$q_{\theta}(y_{n}\mid x_{n-1})=h_{\theta}(y_{n})$, some arbitrary positive
function]. Further, one recovers the bootstrap filter by taking
$q_{\theta}(x_{n}\mid y_{n},x_{n-1})=f_{\theta}(x_{n}\mid x_{n-1})$.
This is an important special case, as some complex models are such
that one may sample from $f_{\theta}(x_{n}\mid x_{n-1})$, but not compute
the corresponding density; in such a case the bootstrap filter is
the only implementable algorithm. For models such that the density
$f_{\theta}(x_{n}\mid x_{n-1})$ is tractable, \cite{Pitt99} recommend
selecting $q_{\theta} (x_{n}\mid y_{n},x_{n-1} )=p_{\theta
} (x_{n}\mid y_{n},x_{n-1} )$
and $q_{\theta} (y_{n}\mid x_{n-1} )=p_{\theta}
(y_{n}\mid x_{n-1} )$
when these quantities are tractable, and using approximations of these
quantities in scenarios when they are not. The intuition for these
recommendations is that this should make the weight function (\ref
{eqimportanceweights})
nearly constant.

The computational complexity of Algorithm \ref{algAPF} is
${\normalcolor\mathcal{O}} (N )$
per time step; in particular, see, for example, \cite{Douc00}, page~201, for a
${\normalcolor\mathcal{O}} (N )$ implementation of the
resampling step. At time $n$, the approximations of $p_{\theta}
( x_{0\dvtx n}\mid y_{0\dvtx n} )$
and $p_{\theta} ( y_{n}\mid y_{0\dvtx n-1} )$ presented
earlier in (\ref{eqfilteringdensityjoint}) and (\ref
{eqrecursivelikeliDecomp}),
respectively, are given by
%
\begin{eqnarray}
\hat{p}_{\theta} ( dx_{0\dvtx n}\mid y_{0\dvtx n} ) & =&\sum
_{i=1}^{N}W_{n}^{i}
\delta_{X_{0\dvtx n}^{i}} (dx_{0\dvtx n} ),\label{eqSMCfullPosterior}
\\
\hat{p}_{\theta} ( y_{n}\mid y_{0\dvtx n-1} ) & =& \Biggl(
\frac{1}{N}\sum_{i=1}^{N}w_{n}
\bigl(X_{n-1\dvtx n}^{i} \bigr) \Biggr)
\nonumber\\[-8pt]\\[-8pt]\nonumber
&&{}\cdot \Biggl(\sum_{i=1}^{N}W_{n-1}^{i}q_{\theta}
\bigl(y_{n}\mid X_{n-1}^{i} \bigr) \Biggr),\hspace*{-25pt}
\end{eqnarray}
where $W_{n}^{i}\propto w_{n} (X_{n-1\dvtx n}^{i} ), \sum_{i=1}^{N}W_{n}^{i}=1$
and $\hat{p}_{\theta} (y_{0} )=\frac{1}{N}\sum_{i=1}^{N}w_{0}
(X_{0}^{i} )$.
In practice, one uses (\ref{eqSMCfullPosterior}) mostly to obtain
approximations of posterior moments
\[
\sum_{i=1}^{N}W_{n}^{i}
\varphi\bigl(X_{0\dvtx n}^{i}\bigr)\approx\mathbb{E} \bigl[
\varphi(X_{0\dvtx n})\mid y_{0\dvtx n} \bigr],
\]
but expressing particle filtering as a method for approximating distributions
(rather than moments) turns out to be a more convenient formalization.
The likelihood~(\ref{eqlikelihoodDecomp}) is then estimated through
%
\begin{equation}
\hat{p}_{\theta} (y_{0\dvtx n} )=\hat{p}_{\theta
}
(y_{0} )\prod
_{k=1}^{n}
\hat{p}_{\theta
} ( y_{k}\mid y_{0\dvtx k-1}
).\label{eqSMClikelihood}
\end{equation}
The resampling procedure is introduced to replicate particles with
high weights and discard particles with low weights. It serves to
focus the computational efforts on the ``promising'' regions
of the state space. We have presented above the simplest resampling
scheme. Lower variance resampling schemes have been proposed in \cite
{Kita96a,Liu98}, as well as more advanced particle algorithms with better
overall performance, for example, the Resample--Move algorithm \cite{Gilks01}.
For the sake of simplicity, we have also presented a version of the
algorithm that operates resampling at every iteration~$n$. It may
be more efficient to trigger resampling only when a certain criterion
regarding the degeneracy of the weights is met; see \cite{Douc00} and
\cite{LiuBook}, pages~35 and~74.

\subsubsection{Convergence results}\label{secconvergenceparticle}

Many sharp convergen\-ce results are available for particle methods
\cite{delmoral2004}. A selection of these results that gives useful
insights on the difficulties of estimating static parameters with
particle methods is presented below.

Under minor regularity assumptions, one can show that for any $n\geq0$,
$N>1$ and any bounded test function $\varphi_{n}\dvtx \mathcal
{X}^{n+1}\rightarrow{}[-1,1]$,
there exist constants $A_{\theta,n,p}<\infty$ such that for any
$p\geq1$
%
\begin{eqnarray}\label{eqweakresult1}
\quad && \mathbb{E} \biggl[\biggl\llvert\int\varphi_{n}(x_{0\dvtx n})\nonumber
\\
&&\hspace*{25pt}{}\cdot \bigl\{ \hat{p}_{\theta} ( dx_{0\dvtx n}\mid y_{0\dvtx n}
)-p_{\theta} ( dx_{0\dvtx n}\mid y_{0\dvtx n} ) \bigr\} \biggr
\rrvert^{p} \biggr]
\\
&&\quad \leq\frac{A_{\theta,n,p}}{N^{p/2}},\nonumber
\end{eqnarray}
where the expectation is with respect to the law of the particle filter.
In addition, for more general classes of functions, we can obtain
for any fixed $n$ a Central Limit Theorem (CLT) as $N\rightarrow
+\infty$
(\cite{chopin2004} and \cite{delmoral2004}, Proposition 9.4.2). Such
results are reassuring but weak, as they reveal nothing regarding
long-time behavior. For instance, without further restrictions on the class
of functions $\varphi_{n}$ and the state-space model, $A_{\theta,n,p}$
typically grows exponentially with $n$. This is intuitively not surprising,
as the dimension of the target density $p_{\theta} (x_{0\dvtx n}\mid
y_{0\dvtx n} )$
is increasing with $n$. Moreover, the successive resampling steps
lead to a depletion of the particle population; $p_{\theta}
( x_{0\dvtx m}\mid y_{0\dvtx n} )$
will eventually be approximated by a single unique particle as $n-m$
\mbox{increases}. This is referred to as the \emph{degeneracy} problem in
the literature (\cite{cappe2005}, Figure~8.4, page 282). This is a fundamental
weakness of particle methods:  given a fixed number of particles $N$,
it is impossible to approximate $p_{\theta}
(x_{0\dvtx n}\mid y_{0\dvtx n} )$
accurately when $n$ is large enough.

Fortunately, it is also possible to establish much more positive results.
Many state-space models possess the so-called \emph{exponential forgetting}
property (\cite{delmoral2004}, Chapter~4). This property states that
for any $x_{0},x_{0}^{\prime}\in\mathcal{X}$ and data $y_{0\dvtx n}$,
there exist constants $B_{\theta}<\infty$ and
$\lambda\in[0,1)$ such that
%
\begin{eqnarray}\label{eqergod}
\quad&& \bigl\llVert p_{\theta} ( dx_{n}\mid y_{1\dvtx n},x_{0}
)-p_{\theta} \bigl( dx_{n}\mid y_{1\dvtx n},x_{0}^{\prime}
\bigr)\bigr\rrVert_{\mathrm{TV}}
\nonumber\\[-8pt]\\[-8pt]\nonumber
&&\quad \leq B_{\theta}\lambda^{n},
\end{eqnarray}
where $\llVert\cdot\rrVert_{\mathrm{TV}}$ is the total variation
distance, that is, the optimal filter forgets exponentially fast its
initial condition. This property is typically satisfied when the signal
process $ \{ X_{n} \} _{n\geq0}$ is a uniformly ergodic
Markov chain and the observations $ \{ Y_{n} \} _{n\geq0}$
are not too informative (\cite{delmoral2004}, Chapter~4), or when $
\{ Y_{n} \} _{n\geq0}$
are informative enough that it effectively restricts the hidden state
to a bounded region around it \cite{outjanerubenthal}. Weaker conditions
can be found in \cite{doucmoulinritov,whiteley2011}. When exponential
forgetting holds, it is possible to establish much stronger uniform-in-time
convergence results \emph{for functions $\varphi_{n}$ that depend
only on recent states}. Specifically, for an integer $L>0$ and any
bounded test function $\Psi_{L}\dvtx \mathcal{X}^{L}\rightarrow{}[-1,1]$,
there exist constants $C_{\theta,L,p}<\infty$ such that for any
$p\geq1$,
$n\geq L-1$,
%
\begin{eqnarray}\label{eqLp}
\quad && \mathbb{E} \biggl[\biggl\llvert\int_{\mathcal{X}^{L}}\Psi
(x_{n-L+1\dvtx n})\Delta_{\theta,n} (dx_{n-L+1\dvtx n} )\biggr\rrvert
^{p} \biggr]
\nonumber\\[-8pt]\\[-8pt]\nonumber
&&\quad \leq\frac{C_{\theta,L,p}}{N^{p/2}},
\end{eqnarray}
where
%
\begin{eqnarray}\label{eq3}
&& \Delta_{\theta,n} (dx_{n-L+1\dvtx n} )\nonumber
\\
&&\quad =\int_{x_{0\dvtx n-L}\in
\mathcal{X}^{n-L+1}} \bigl\{
\hat{p}_{\theta} (dx_{0\dvtx n}\mid y_{0\dvtx n})
\\
&&\hspace*{91pt}{} -p_{\theta} (dx_{0\dvtx n}\mid y_{0\dvtx n} ) \bigr\}.\hspace*{-10pt}\nonumber
\end{eqnarray}
This result explains why particle filtering is an effective computational
tool in many applications such as tracking, where one is only interested
in $p_{\theta} ( x_{n-L+1\dvtx n}\mid y_{0\dvtx n} )$,
as the approximation error is uniformly bounded over time.

Similar positive results hold for $\hat{p}_{\theta}
(y_{0\dvtx n} )$.
This estimate is unbiased for any $N\geq1$ (\cite{delmoral2004}, Theorem 7.4.2, page~239),
and, under assumption (\ref{eqergod}), the \emph{relative} variance
of the likelihood estimate $\hat{p}_{\theta} (y_{0\dvtx n} )$,
that is the variance of the ratio $\hat{p}_{\theta}
(y_{0\dvtx n} )/p_{\theta} (y_{0\dvtx n} )$,
is bounded above by $D_{\theta}n/N$ \cite{cerou2010,whiteley2011}.
This is a great improvement over the exponential increase with $n$
that holds for standard importance sampling techniques; see, for instance,
\cite{doucet2011}. However, the constants $C_{\theta,L,p}$ and
$D_{\theta}$
are typically exponential in $n_{x}$, the dimension of the state
vector $X_{n}$. We note that nonstandard particle methods designed
to minimize the variance of the estimate of $p_{\theta}
(y_{0\dvtx n} )$
have recently been proposed \cite{whiteley2014}.

Finally, we recall the theoretical properties of particles estimates
of the following so-called smoothed additive functional (\cite{cappe2005}, Section~8.3 and
\cite{olsson2008}),
%
\begin{eqnarray}\label{eqexpectationadditivefunctionals}
\mathcal{S}_{n}^{\theta} &=& \int_{\mathcal{X}^{n+1}} \Biggl\{
\sum_{k=1}^{n}s_{k}
(x_{k-1\dvtx k} ) \Biggr\}
\nonumber\\[-8pt]\\[-8pt]\nonumber
&&\hspace*{25pt}{}\cdot  p_{\theta} ( x_{0\dvtx n}\mid y_{0\dvtx n} )\,dx_{0\dvtx n}.
\end{eqnarray}
Such quantities are critical when implementing ML parameter estimation
procedures; see Section~\ref{secMLestimation}. If we substitute
$\hat{p}_{\theta} ( dx_{0\dvtx n}\mid y_{0\dvtx n} )$
to $p_{\theta} ( x_{0\dvtx n}\mid y_{0\dvtx n} )\,dx_{0\dvtx n}$
to approximate $\mathcal{S}_{n}^{\theta}$, then we obtain an estimate
$\widehat{\mathcal{S}}_{n}^{\theta}$ which can be computed recursively
in time; see, for example, \cite{cappe2005}, Section~8.3. For the remainder
of this paper we will refer to this approximation as the \emph{path
space} approximation. Even when (\ref{eqergod}) holds, there exists
$0<F_{\theta},G_{\theta}<\infty$ such that the asymptotic bias \cite
{delmoral2004}
and variance \cite{poyadjis2009} satisfy
%
\begin{equation}
\qquad\bigl\llvert\mathbb{E} \bigl(\widehat{\mathcal{S}}_{n}^{\theta}
\bigr)-\mathcal{S}_{n}^{\theta}\bigr\rrvert\leq F_{\theta}
\frac{n}{N},\quad \mathbb{ V} \bigl(\widehat{\mathcal{S}}_{n}^{\theta}
\bigr)\geq G_{\theta}\frac{n^{2}}{N}\label{eqsufficientstatsdegrade}
\end{equation}
for $s_{p}\dvtx \mathcal{X}^{2}\rightarrow{}[-1,1]$ where the variance
is w.r.t. the law of the particle filter. The fact that the variance
grows at least quadratically in time follows from the degeneracy problem
and makes $\widehat{\mathcal{S}}_{n}^{\theta}$ unsuitable for some
on-line likelihood based parameter estimation schemes discussed in
Section~\ref{secMLestimation}.

\section{Smoothing}\label{secsmoothingandparticle}

In this section the parameter $\theta$ is still assumed known and
we focus on smoothing, that is, the problem of estimating the latent
variables $X_{0\dvtx T}$ given a fixed batch of observations $y_{0\dvtx T}$.
Smoothing for a fixed parameter $\theta$ is at the core of the two
main particle ML parameter inference techniques described in
Section~\ref{secMLestimation}, as these procedures require computing smoothed
additive functionals of the form (\ref{eqexpectationadditivefunctionals}).
Clearly, one could unfold the recursion (\ref{eqjointupdate}) from
$n=0$ to $n=T$ to obtain $p_{\theta} ( x_{0\dvtx T}\mid
y_{0\dvtx T} )$.
However, as pointed out in the previous section, the path space approximation
(\ref{eqSMCfullPosterior}) suffers from the degeneracy problem and
yields potentially high variance estimates of (\ref
{eqexpectationadditivefunctionals})
as (\ref{eqsufficientstatsdegrade}) holds. This has motivated the
development of alternative particle approaches to approximate
$p_{\theta} ( x_{0\dvtx T}\mid y_{0\dvtx T} )$
and its marginals.

\subsection{Fixed-lag Approximation}
\label{subFixedLagapproxparticle}

For state-space models with ``good'' forgetting
properties [e.g., (\ref{eqergod})], we have
%
\begin{equation}
p_{\theta} ( x_{0\dvtx n}\mid y_{0\dvtx T} )\approx
p_{\theta} ( x_{0\dvtx n}\mid y_{0\dvtx  (n+L
)\wedge T} )\label{eqforgetting}
\end{equation}
for $L$ large enough, that is, observations collected at times $k>n+L$
do not bring any significant additional information about $X_{0\dvtx n}$.
In particular, when having to evaluate $\mathcal{S}_{T}^{\theta}$
of the form (\ref{eqexpectationadditivefunctionals}) we can approximate
the expectation of $s_{n} (x_{n-1\dvtx n} )$ w.r.t. $p_{\theta
} ( x_{n-1\dvtx n}\mid y_{0\dvtx T} )$
by its expectation w.r.t. $p_{\theta} ( x_{n-1\dvtx n}\mid y_{0\dvtx  (n+L )\wedge
T} )$.

Algorithmically, a particle implementation of (\ref{eqforgetting})
means not resampling the components $X_{0\dvtx n}^{i}$ of the particles
$X_{0\dvtx k}^{i}$ obtained by particle filtering at times $k>n+L$. This
was first suggested in \cite{kitagawa2001} and used in~\cite{cappe2005}, Section 8.3, and \cite{olsson2008}. This algorithm is simple to
implement, but the
main practical problem is the choice of $L$. If taken too small,
then $p_{\theta} ( x_{0\dvtx n}\mid y_{0\dvtx
(n+L )\wedge T)} )$
is a poor approximation of $p_{\theta} ( x_{0\dvtx n}\mid y_{0\dvtx T} )$.
If taken too large, the degeneracy remains substantial. Moreover,
even as $N\rightarrow\infty$, this particle approximation will
have a nonvanishing bias since $p_{\theta} ( x_{0\dvtx n}\mid y_{0\dvtx T} )\neq
p_{\theta} ( x_{0\dvtx n}\mid
y_{0\dvtx  (n+L )\wedge T} )$.

\subsection{Forward--Backward Smoothing}
\label{subForward-backwardsmoothing}

\subsubsection{Principle}

The joint smoothing density\break 
$p_{\theta} ( x_{0\dvtx T}\mid y_{0\dvtx T} )$ can be
expressed as a function of the filtering densities $ \{ p_{\theta
} ( x_{n}\mid y_{0\dvtx n} ) \} _{n=0}^{T}$
using the following key decomposition:
%
\begin{eqnarray}\label{eqjointdecomposition}
&& p_{\theta} ( x_{0\dvtx T}\mid y_{0\dvtx T} )
\nonumber\\[-8pt]\\[-10pt]\nonumber
&&\quad =p_{\theta
} (x_{T}\mid y_{0\dvtx T} )\prod
_{n=0}^{T-1}p_{\theta}
( x_{n}\mid y_{0\dvtx n},x_{n+1} ),
\end{eqnarray}
where $p_{\theta} ( x_{n}\mid y_{0\dvtx n},x_{n+1} )$
is a backward (in time) Mar\-kov transition density given by
%
\begin{equation}
\qquad p_{\theta} ( x_{n}\mid y_{0\dvtx n},x_{n+1} )=
\frac{f_{\theta} ( x_{n+1}\mid x_{n}
)p_{\theta} ( x_{n}\mid y_{0\dvtx n} )}{p_{\theta
} ( x_{n+1}\mid y_{0\dvtx n} )}.\label{eqbackwardkernel}
\end{equation}
A backward in time recursion for $ \{ p_{\theta} (x_{n}\mid y_{0\dvtx T} )
\} _{n=0}^{T}$
follows by integrating out $x_{0\dvtx n-1}$ and $x_{n+1\dvtx T}$ in (\ref
{eqjointdecomposition})
while applying (\ref{eqbackwardkernel}),
%
\begin{eqnarray}\label{eqmarginalforwardbackward}
\qquad && p_{\theta} ( x_{n}\mid y_{0\dvtx T} )\nonumber
\\
&&\quad =p_{\theta
} (x_{n}\mid y_{0\dvtx n} )
\\
&&\qquad{} \cdot \int\frac{f_{\theta
} ( x_{n+1}\mid x_{n} )p_{\theta} (x_{n+1}\mid y_{0\dvtx T} )}{p_{\theta}
(x_{n+1}\mid y_{0\dvtx n} )}\,dx_{n+1}.\nonumber
\end{eqnarray}
This is referred to as forward--backward smoothing, as a forward pass
yields $ \{ p_{\theta} ( x_{n}\mid
y_{0\dvtx n} ) \} _{n=0}^{T}$
which can be used in a backward pass to obtain $ \{ p_{\theta
} ( x_{n}\mid\break  y_{0\dvtx T} ) \} _{n=0}^{T}$.
Combined to $ \{ p_{\theta} ( x_{n}\mid
y_{0\dvtx n},x_{n+1} ) \} _{n=0}^{T-1}$,
this allows us to obtain $\mathcal{S}_{T}^{\theta}$. An alternative
to these forward--backward procedures is the generalized two-filter
formula \cite{Briers09}.

\subsubsection{Particle implementation}

The decomposition (\ref{eqjointdecomposition}) suggests that it
is possible to sample approximately from $p_{\theta} (x_{0\dvtx T}\mid
y_{0\dvtx T} )$
by running a particle filter from time $n=0$ to $T$, storing the
approximate filtering distributions $ \{ \hat{p}_{\theta
} ( dx_{n}\mid y_{0\dvtx n} ) \} _{n=0}^{T}$,
that is, the marginals of (\ref{eqSMCfullPosterior}), then sampling
$X_{T}\sim\hat{p}_{\theta} ( dx_{T}\mid
y_{0\dvtx T} )$
and for $n=T-1,T-2,\ldots,0$ sampling $X_{n}\sim\hat{p}_{\theta
} ( dx_{n}\mid y_{0\dvtx n},X_{n+1} )$
where this distribution is obtained by substituting $\hat{p}_{\theta} (
dx_{n}\mid y_{0\dvtx n} )$
for $p_{\theta} ( dx_{n}\mid y_{0\dvtx n} )$ in
(\ref{eqbackwardkernel}):
%
\begin{eqnarray}\label{eqbackwardparticle-new}
&& \hat{p}_{\theta} ( dx_{n}\mid y_{0\dvtx n},X_{n+1})
\nonumber\\[-8pt]\\[-8pt]\nonumber
&&\quad =\frac{\sum_{i=1}^{N}W_{n}^{i}f_{\theta
}(X_{n+1}|X_{n}^{i})\delta_{X_{n}^{i}} (dx_{n} )}{\sum
_{i=1}^{N}W_{n}^{i}f_{\theta}(X_{n+1}|X_{n}^{i})}.
\end{eqnarray}
This Forward Filtering Backward Sampling (FFBSa) procedure was proposed
in \cite{godsill}. It requires $\mathcal{O} (N (T+1
) )$
operations to generate a single path $X_{0\dvtx T}$, as sampling from (\ref
{eqbackwardparticle-new})
costs $\mathcal{O} (N )$ operations. However, as noted in
\cite{douc09}, it is possible to sample using rejection from an alternative
approximation of $p_{\theta} ( x_{n}\mid
y_{0\dvtx n},X_{n+1} )$
in $\mathcal{O} (1 )$ operations if we use an
unweighted particle approximation of $p_{\theta} (x_{n}\mid y_{0\dvtx n} )$
in~(\ref{eqbackwardkernel}) and if the transition prior satisfies
$f_{\theta}(x^{\prime}\mid x)\leq C<\infty$. Hence, with this approach,
sampling a path $X_{0\dvtx T}$ costs, on average, only $\mathcal{O}
(T+1 )$
operations. A~related rejection technique was proposed in \cite{hurzeler1998}.
In practice, one may generate $N$ such trajectories to compute Monte
Carlo averages that approximate smoothing expectations $\mathbb
{E} [\varphi(X_{0\dvtx T} )\mid y_{0\dvtx T} ]$.
In that scenario, the first approach costs $\mathcal{O}
(N^{2} (T+1 ) )$,
while the second approach costs $\mathcal{O} (N (T+1
) )$ on average. In some applications, the
rejection sampling procedure can be computationally costly as the
acceptance probability can be very small for some particles; see,
for example, Section~4.3 in \cite{olsson2014} for empirical results.
This has
motivated the development of hybrid procedures combining FFBSa and
rejection sampling \cite{taghavi2013}.

We can also directly approximate the marginals $ \{ p_{\theta
} ( x_{n}\mid y_{0\dvtx T} ) \} _{n=0}^{T}$.
Assuming we have an approximation $\bar{p}_{\theta} (dx_{n+1}\mid
y_{0\dvtx T} )=\sum_{i=1}^{N}W_{n+1\mid T}^{i}\delta_{X_{n+1}^{i}}(dx_{n+1})$
where $W_{ T\mid T}^{i}=W_{T}^{i}$, then by using (\ref
{eqmarginalforwardbackward})
and (\ref{eqbackwardparticle-new}), we obtain the approximation
$\bar{p}_{\theta} ( dx_{n}\mid y_{0\dvtx T}
)= \break \sum_{i=1}^{N}W_{ n\mid T}^{i}\delta_{X_{n}^{i}}(dx_{n})$
with
%
\begin{equation}
\quad W_{n\mid T}^{i}=W_{n}^{i}\times\sum
_{j=1}^{N}\frac
{W_{n+1\mid T}^{j}f_{\theta} (X_{n+1}^{j}\mid X_{n}^{i} )}{\sum
_{l=1}^{N}W_{n}^{l}f_{\theta} (X_{n+1}^{j}\mid X_{n}^{l}
)}.\label{eqbackwardweights-new}
\end{equation}
This Forward Filtering Backward Smoothing\break (\mbox{FFBSm}, where ``m'' stands
for ``marginal'') procedure requires $\mathcal{O} (N^{2}
(T+1 ) )$
operations to approximate $ \{ p_{\theta} (x_{n}\mid y_{0\dvtx T} ) \} _{n=0}^{T}$
instead of $\mathcal{O} (N (T+1 ) )$ for the path
space and fixed-lag methods. However, this high computational complexity
of forward--backward estimates can be reduced using fast computational
methods \cite{Klass2005}. Particle approximations of generalized
two-filter smoothing procedures have also been proposed in \cite
{Briers09,fearnhead2008}.

\subsection{Forward Smoothing}
\label{subForward-only-smoothing}

\subsubsection{Principle}

Whenever we are interested in computing the sequence
$ \{ \mathcal{S}_{n}^{\theta} \} _{n\geq0}$ recursively
in time, the forward--backward
procedure described above is cumbersome, as it requires performing
a new backward pass with $n+1$ steps at time $n$. An important but
not well-known result is that it is possible to implement exactly
the forward--backward procedure using only a forward procedure. This
result is at the core of \cite{Elliott1996}, but its exposition relies
on tools which are nonstandard for statisticians. We follow here
the simpler derivation proposed in \cite{delmoralforward,delmoral2009} which simply
consists of rewriting (\ref{eqexpectationadditivefunctionals}) as
%
\begin{equation}
\mathcal{S}_{n}^{\theta}=\int V_{n}^{\theta}
(x_{n} )p_{\theta} ( x_{n}\mid y_{0\dvtx n}
)\,dx_{n},\label{eqadditivesmoothfunctionalsasfunctionofT}
\end{equation}
where
%
\begin{eqnarray}\label{eqdefinitionVn}
\quad V_{n}^{\theta} (x_{n} )&:=&\int\Biggl\{ \sum
_{k=1}^{n}s_{k} (x_{k-1\dvtx k} ) \Biggr\}
\nonumber\\[-8pt]\\[-8pt]\nonumber
&&\hspace*{8pt}{}\cdot p_{\theta} ( x_{0\dvtx n-1}\mid y_{0\dvtx n-1},x_{n}
)\,dx_{0\dvtx n-1}.
\end{eqnarray}
It can be easily checked using (\ref{eqjointdecomposition}) that
$V_{n}^{\theta} (x_{n} )$ satisfies the following forward
recursion for $n\geq0$:
%
\begin{eqnarray}\label{eqrecursionadditivefunctionalVn}
V_{n+1}^{\theta} (x_{n+1} ) &=& \int\bigl\{
V_{n}^{\theta
} (x_{n} )+s_{n+1}
(x_{n\dvtx n+1} ) \bigr\}
\nonumber\\[-8pt]\\[-10pt]\nonumber
&&\hspace*{9pt}{}\cdot p_{\theta} ( x_{n}\mid y_{0\dvtx n},x_{n+1} )\,dx_{n},
\end{eqnarray}
with $V_{0}^{\theta} (x_{0} )=0$ and where $p_{\theta
} ( x_{n}\mid y_{0\dvtx n},x_{n+1} )$
is given by (\ref{eqbackwardkernel}). In practice, we shall approximate
the function $V_{n}^{\theta}$ on a certain grid of values $x_{n}$,
as explained in the next section.

\subsubsection{Particle implementation}

We can easily provide a particle approximation of the forward smoothing
recursion. Assume you have access to approximations $ \{ \widehat
{V}_{n}^{\theta} (X_{n}^{i} ) \} $
of $ \{ V_{n}^{\theta} (X_{n}^{i} ) \} $ at time
$n$, where $\hat{p}_{\theta} ( dx_{n}\mid
y_{0\dvtx n} )=\sum_{i=1}^{N}W_{n}^{i}\delta_{X_{n}^{i}}
(dx_{n} )$.
Then when updating our particle\vspace*{1pt} filter to obtain $\hat{p}_{\theta
} ( dx_{n+1}\mid y_{0\dvtx n+1} )=\sum_{i=1}^{N}W_{n+1}^{i}\delta
_{X_{n+1}^{i}} (dx_{n+1} )$,
we can directly compute the particle approximations $ \{ \widehat
{V}_{n+1}^{\theta} (X_{n+1}^{i} ) \} $
by\vspace*{1pt} plugging (\ref{eqbackwardparticle-new}) and $\hat{p}_{\theta
} ( dx_{n}\mid y_{0\dvtx n} )$
in (\ref{eqadditivesmoothfunctionalsasfunctionofT})--(\ref
{eqrecursionadditivefunctionalVn})
to obtain
%
\begin{eqnarray}
\widehat{V}_{n+1}^{\theta} \bigl(X_{n+1}^{i}\bigr)
&=& \Biggl(\sum_{j=1}^{N}W_{n}^{j}f_{\theta} \bigl(X_{n+1}^{i}\mid
X_{n}^{j} \bigr)\nonumber
\\
\label{eqTapproximation-new} &&\hspace*{22pt}{}\cdot
\bigl\{ \widehat{V}_{n}^{\theta} \bigl(X_{n}^{j} \bigr)+s_{n+1}
\bigl(X_{n}^{j},X_{n+1}^{i} \bigr) \bigr\} \Biggr)
\\
&&{} \Big/ \Biggl(\sum_{j=1}^{N}W_{n}^{j}f_{\theta}
\bigl(X_{n+1}^{i}\mid X_{n}^{j}
\bigr)\Biggr),\nonumber
\\
\widehat{\mathcal{S}}_{n}^{\theta}  &=& \sum
_{i=1}^{N}W_{n}^{i}
\widehat{V}_{n}^{\theta} \bigl(X_{n}^{i}
\bigr).\label{eqSMCapproxadditivefunctionals-1}
\end{eqnarray}
This approach requires $\mathcal{O} (N^{2} (n+1
) )$
operations to compute $\widehat{\mathcal{S}}_{n}^{\theta}$ at iteration
$n$. A variation over this idea recently proposed in \cite{olsson2014}
and \cite{westerbornOlsson2014}
consists of approximating $V_{n+1}^{\theta} (X_{n+1}^{i} )$
by sampling $X_{n}^{i,j}\sim\hat{p}_{\theta} ( dx_{n}\mid
y_{0\dvtx n}, X_{n+1}^{i} )$
for $j=1,\ldots,K$ to obtain
%
\begin{eqnarray}\label{eqOlssonMCapproximation}
&& \widehat{V}_{n+1}^{\theta} \bigl(X_{n+1}^{i}
\bigr)
\nonumber\\[-8pt]\\[-8pt]\nonumber
&&\quad =\frac{1}{K}\sum_{j=1}^{K}
\bigl\{ \widehat{V}_{n}^{\theta} \bigl(X_{n}^{i,j}
\bigr)+s_{n+1} \bigl(X_{n}^{i,j},X_{n+1}^{i}
\bigr) \bigr\}.\hspace*{-20pt}
\end{eqnarray}
When\vspace*{1pt} it is possible to sample from $\hat{p}_{\theta} ( dx_{n}\mid
y_{0\dvtx n},X_{n+1}^{i} )$
in $\mathcal{O} (1 )$ operations using rejection sampling,
(\ref{eqOlssonMCapproximation})
provides a Monte Carlo approximation to (\ref{eqTapproximation-new})
of overall complexity $\mathcal{O} (NK )$.

\subsection{Convergence Results for Particle Smoothing}

Empirically, for a fixed number of particles, these smoothing procedures
perform significantly much better than the naive path space approach
to smoothing (i.e., simply propagating forward the complete state
trajectory within a particle filtering algorithm). Many theoretical
results validating these empirical findings have been established
under assumption (\ref{eqergod}) and additional regularity
assumptions. The particle estimate
of $\mathcal{S}_{n}^{\theta}$ based on the fixed-lag approximation
(\ref{eqforgetting}) has an asymptotic variance in $n/N$ with a
nonvanishing (as $N\rightarrow\infty$) bias proportional to $n$
and a constant decreasing exponentially fast with $L$ \cite{olsson2008}.
In \mbox{\cite{delmoral2009,delmoralforward,douc09}},
it is shown that when (\ref{eqergod}) holds, there exists $0<F_{\theta
}$, $H_{\theta}<\infty$
such that\vspace*{1pt} the asymptotic bias and variance of the particle estimate
of $\mathcal{S}_{n}^{\theta}$ computed using the forward--backward
procedures satisfy
%
\begin{equation}
\bigl\llvert\mathbb{E} \bigl(\widehat{\mathcal{S}}_{n}^{\theta}
\bigr)-\mathcal{S}_{n}^{\theta}\bigr\rrvert\leq F_{\theta}
\frac{n}{N},\qquad \mathbb{ V} \bigl(\widehat{\mathcal{S}}_{n}^{\theta}
\bigr)\leq H_{\theta}\frac{n}{N}.\hspace*{-30pt}\label{eqsufficientstatsFB}
\end{equation}
The bias for the path space and forward--backward estimators of
$\mathcal{S}_{n}^{\theta}$
are actually equal \cite{delmoralforward}. Recently, it has also been
established in \cite{olsson2014} that, under similar regularity
assumptions, the
estimate obtained through~(\ref{eqOlssonMCapproximation}) also admits an asymptotic
variance in $n/N$ whenever $K\geq2$.

\section{Maximum Likelihood Parameter Estimation}\label{secMLestimation}

We describe in this section how the particle filtering and smoothing
techniques introduced in Sections~\ref{secfilteringandparticle}
and \ref{secsmoothingandparticle} can be used to implement maximum
likelihood parameter estimation techniques.

\subsection{Off-Line Methods}

\label{secofflineML} We recall that $\ell_{T} (\theta)$
denote the log-likelihood function associated to data $y_{0\dvtx T}$ introduced
in Section~\ref{seccomputationalissues}. So as to maximize $\ell
_{T} (\theta)$,
one can rely on standard nonlinear optimization methods, for example, using
quasi-Newton or gradient-ascent techniques. We will limit ourselves
to these approaches even if they are sensitive to initialization and
might get trapped in a local maximum.

\subsubsection{Likelihood function evaluation}

We have seen in Section~\ref{secfilteringandparticle} that $\ell
_{T} (\theta)$
can be approximated using particle methods, for any fixed $\theta\in
\Theta$.
One may wish then to treat ML estimation as an optimization problem
using Monte Carlo evaluations of $\ell_{T} (\theta)$. When
optimizing a function calculated with a Monte Carlo error, a popular
strategy is to make the evaluated function continuous by using common
random numbers over different evaluations to ease the optimization.
Unfortunately, this strategy is not helpful in the particle context.
Indeed, in the resampling stage, particles $\{\overline{X}_{n}^{i}\}_{i=1}^{N}$
are resampled according to the distribution $\sum_{i=1}^{N}\overline
{W}_{n+1}^{i}\delta_{X_{n}^{i}} (dx_{n} )$
which admits a piecewise constant and hence discontinuous cumulative
distribution function (c.d.f.). A small change in $\theta$ will cause
a small change in the importance weights $\{\overline{W}_{n+1}^{i}\}_{i=1}^{N}$
and this will potentially generate a different set of resampled particles.
As a result, the log-likelihood function estimate will not be continuous
in $\theta$ even if $\ell_{T} (\theta)$ is continuous.

To bypass this problem, an importance sampling method was introduced
in \cite{Hurzeler2001}, but it has computational complexity $\mathcal
{O} (N^{2} (T+1 ) )$
and only provides low variance estimates in the neighborhood of a
suitably preselected parameter value. In the restricted scenario where
$\mathcal{X}\subseteq\mathbb{R}$, an elegant solution to the discontinuity
problem was proposed in \cite{Pitt02}. The method uses common random
numbers and introduces a ``continuous'' version
of the resampling step by finding a permutation~$\sigma$ such that
$X_{n}^{\sigma(1 )}\leq X_{n}^{\sigma(2
)}\leq\cdots\leq X_{n}^{\sigma(N )}$
and defining a piecewise linear approximation of the resulting c.d.f.
from which particles are resampled, that is,
%
\begin{eqnarray}
F_{n} (x )= \Biggl(\sum_{i=1}^{k-1}
\overline{W}_{n+1}^{\sigma(i )} \Biggr)+\overline{W}_{n+1}^{\sigma
(k )}
\frac{x-X_{n}^{\sigma(k-1
)}}{X_{n}^{\sigma(k )}-X_{n}^{\sigma(k-1
)}},\nonumber
\\
\eqntext{X_{n}^{\sigma(k-1 )}\leq x\leq
X_{n}^{\sigma (k )}.}
\end{eqnarray}
This method requires $\mathcal{O} (N (T+1 )\log
N )$
operations due to the sorting of the particles, but the resulting continuous
estimate of $\ell_{T} (\theta)$ can be maximized using
standard optimization techniques. Extensions to the multivariate case
where $\mathcal{X}\subseteq\mathbb{R}^{n_{x}}$ (with $n_{x}>1$)
have been proposed in \cite{Lee08} and \cite{DeJong2013}. However,
the scheme \cite{Lee08} does not guarantee continuity of the likelihood
function estimate and only provides log-likelihood estimates which
are positively correlated for neighboring values in the parameter
space, whereas the scheme in \cite{DeJong2013} has $\mathcal{O}
(N^{2} )$
computational complexity and relies on a nonstandard particle filtering
scheme.

When $\theta$ is high dimensional, the optimization over the parameter
space may be made more efficient if provided with estimates of the
gradient. This is exploited by the algorithms described in the forthcoming
sections.

\subsubsection{Gradient ascent}

The log-likelihood $\ell_{T} (\theta)$ may be maximized
with the following steepest ascent algorithm:  at iteration $k+1$
%
\begin{equation}
\theta_{k+1}=\theta_{k}+\gamma_{k+1}
\nabla_{\theta}\ell_{T}(\theta)\mid_{\theta=\theta_{k}},\label
{eqbatchgradient}
\end{equation}
where $ \nabla_{\theta}\ell_{T}(\theta)\mid_{\theta
=\theta_{k}}$
is the gradient of $\ell_{T}(\theta)$ w.r.t. $\theta$ evaluated at
$\theta=\theta_{k}$ and $\{\gamma_{k}\}$ is a sequence of positive
real numbers, called the step-size sequence. Typically, $\gamma_{k}$
is determined adaptively at iteration $k$ using a line search or
the popular Barzilai--Borwein alternative. Both schemes guarantee convergence
to a local maximum under weak regularity assumptions; see \cite{yuan2008}
for a survey.

The \emph{score} vector $\nabla_{\theta}\ell_{T} (\theta
)$
can be computed by using Fisher's identity given in (\ref
{eqfisheridentityloglike}).
Given (\ref{eqjointpdfxy}), it is easy to check that the score is
of the form (\ref{eqexpectationadditivefunctionals}). An alternative
to Fisher's identity to compute the score is presented in \cite{coquelin2008},
but this also requires computing an expectation of the form (\ref
{eqexpectationadditivefunctionals}).

These score estimation methods are not applicable in complex scenarios
where it is possible to sample from $f_{\theta}(x^{\prime}\mid x)$, but
the analytical expression of this transition kernel is unavailable
\cite{ionidesbretoking2006}. For those models, a naive approach
is to use a finite difference estimate of the gradient; however, this
might generate too high a variance estimate. An interesting alternative
presented in \cite{Ionides09}, under the name of iterated filtering,
consists of deriving an approximation of $ \nabla_{\theta}\ell
_{T}(\theta)\mid_{\theta=\theta_{k}}$
based on the posterior moments $\{\mathbb{E}(\vartheta
_{n}\mid y_{0\dvtx n}),\mathbb{V}(\vartheta_{n}\mid y_{0\dvtx n})\}_{n=0}^{T}$
of an artificial state-space model with latent Markov process $\{
Z_{n}=(X_{n},\vartheta_{n})\}_{n=0}^{T}$,
%
\begin{equation}
\qquad \vartheta_{n+1}=\vartheta_{n}+\varepsilon_{n+1},\quad
X_{n+1}\sim f_{\vartheta_{n+1}}(\cdot\mid x_{n}),\quad
\end{equation}
and observed process $Y_{n+1}\sim g_{\vartheta_{n+1}}(\cdot\mid x_{n+1})$.
Here $ \{ \varepsilon_{n} \} _{n\geq1}$ is a zero-mean white
noise sequence with variance $\sigma^{2}\Sigma$, $\mathbb
{E}(\vartheta_{n+1}\mid\vartheta_{n})=\vartheta_{n}$,
$\mathbb{E}(\vartheta_{0})=\theta_{k}$, $\mathbb{V}(\vartheta
_{0})=\tau^{2}\Sigma$.
It is shown in \cite{Ionides09} that this approximation improves
as $\sigma^{2},\tau^{2}\rightarrow0$ and $\sigma^{2}/\tau
^{2}\rightarrow0$.
Clearly, as the variance $\sigma^{2}$ of the artificial dynamic noise
$ \{ \varepsilon_{n} \} $ on the $\theta$-component decreases,
it will be necessary to use more particles to approximate $ \nabla
_{\theta}\ell_{T}(\theta)\mid_{\theta=\theta_{k}}$
as the mixing properties of the artificial dynamic model deteriorates.

\subsubsection{Expectation--Maximization}

Gradient ascent algorithms can be numerically unstable as they require
to scale carefully the components of the score vector. The Expectation
Maximization (EM) algorithm is a very popular alternative procedure
for maximizing $\ell_{T}(\theta)$ \cite{Demster77}. At iteration
$k+1$, we set
%
\begin{equation}
\theta_{k+1}=\arg\max_\theta Q(\theta_{k},
\theta),\label{eqmaximizationEM}
\end{equation}
where
%
\begin{eqnarray}\label{eqQfunction}
Q(\theta_{k},\theta) &=& \int\log p_{\theta
}(x_{0\dvtx T},y_{0\dvtx T})
\nonumber\\[-8pt]\\[-8pt]\nonumber
&&\hspace*{9pt}{}\cdot p_{\theta _{k}}(x_{0\dvtx T}\mid y_{0\dvtx T})\,dx_{0\dvtx T}.
\end{eqnarray}
The sequence $\{\ell_{T}(\theta_{k})\}_{k\geq0}$ generated by this
algorithm is nondecreasing. The EM is usually favored by practitioners
whenever it is applicable, as it is numerically more stable than gradient
techniques.

In terms of implementation, the EM consists of computing a $n_{s}$-dimensional
summary statistic of the form (\ref{eqexpectationadditivefunctionals})
when $p_{\theta}(x_{0\dvtx T},y_{0\dvtx T})$ belongs to the exponential family,
and the maximizing argument of $Q(\theta_{k},\theta)$ can be characterized
explicitly through a suitable function $\Lambda\dvtx \mathbb
{R}^{n_{s}}\rightarrow\Theta$,
that is,
%
\begin{equation}
\theta_{k+1}=\Lambda\bigl(T^{-1}\mathcal{S}_{T}^{\theta_{k}}
\bigr).\label{eqmaximiEM}
\end{equation}

\subsubsection{Discussion of particle implementations}\label
{subDiscussion-of-particle-batch}

The path space approximation (\ref{eqSMCfullPosterior}) can be used
to approximate the score (\ref{eqfisheridentityloglike}) and the
summary statistics of the EM algorithm at the computational cost of
$\mathcal{O}(N (T+1 ))$; see \cite{cappe2005}, Section 8.3, and \cite{olsson2008,poyadjis2009}. Experimentally, the variance
of the associated estimates increases typically quadratically with
$T$ \cite{poyadjis2009}. To obtain estimates whose variance increases
only typically linearly with $T$ with similar computational cost,
one can use the fixed-lag approximation presented in Section~\ref
{subFixedLagapproxparticle}
or a more recent alternative where the path space method is used, but
the additive functional of interest, which is a sum of terms over
$n=0,\ldots,T$, is approximated by a sum of similar terms which are
now exponentially weighted w.r.t. $n$ \cite{nemeth2013}. These methods
introduce a nonvanishing asymptotic bias difficult to quantify but
appear to perform well in practice.

To improve over the path space method without introducing any such
asymptotic bias, the FFBSm and forward smoothing discussed in
Sections~\ref{subForward-backwardsmoothing} and \ref
{subForward-only-smoothing}
as well as the generalized two-filter smoother have been used \cite
{schon2010,delmoral2009,delmoralforward,poyadjis2009,Briers09}.
Experimentally, the variance of the associated estimates
increases typically linearly with $T$ \cite{poyadjis2009} in agreement
with the theoretical results in \cite
{delmoral2009,delmoralforward,douc09}. However, the computational
complexity of these techniques
is $\mathcal{O}(N^{2} (T+1 ))$. For a fixed computational
complexity of order $\mathcal{O} (N^{2}(T+1) )$, an informal
comparison of the performance of the path space estimate using $N^{2}$
particles and the forward--backward estimate using $N$ particles suggest
that both estimates admit a Mean Square Error (MSE) of order $\mathcal
{O}(N^{-2} (T+1 ))$,
but the MSE of the path space estimate is variance dominated, whereas
the forward--backward estimates are bias dominated. This can be understood
by decomposing the MSE as the sum of the squared bias and the variance
and then substituting appropriately for $N^{2}$ particles in (\ref
{eqsufficientstatsdegrade})
for the path space method and for $N$ particles in (\ref{eqsufficientstatsFB})
for the forward--backward estimates. We confirm experimentally this
fact in Section~\ref{subMaximum-likelihood-methods}.

These experimental results suggest that these particle smoothing estimates
might thus be of limited interest compared to the path based estimates
for ML parameter inference when accounting for computational complexity.
However, this comparison ignores that the $\mathcal{O}(N^{2})$ computational
complexity of these particle smoothing estimates can be reduced to
$\mathcal{O} (N )$ by sampling approximately from
$p_{\theta}(x_{0\dvtx T}\mid y_{0\dvtx T})$
with the FFBSa procedure in Section~\ref{subForward-backwardsmoothing}
or by using fast computational methods \cite{Klass2005}. Related
$\mathcal{O} (N )$ approaches have been developed for generalized
two-filter smoothing \cite{briers2005,fearnhead2008}. When
applicable, these fast computational methods should be favored.

\subsection{On-Line Methods}

For a long observation sequence the computation of the gradient of
$\ell_{T}(\theta)$ can be prohibitive, and moreover, we might have real-time
constraints. An alternative would be a recursive procedure in which
the data is run through once sequentially. If $\theta_{n}$ is the
estimate of the model parameter after the first $n$ observations,
a recursive method would update the estimate to $\theta_{n+1}$ after
receiving the new data $y_{n}$. Several on-line variants of the ML
procedures described earlier are now presented. For these methods
to be justified, it is crucial for the observation process to be ergodic
for the limiting averaged likelihood function $\ell_{T} (\theta
)/T$
to have a well-defined limit $\ell(\theta)$ as
$T\rightarrow+\infty$.

\subsubsection{On-line gradient ascent}

An alternative to gradient ascent is the following parameter update
scheme at time $n\geq0$:
%
\begin{equation}
\qquad \theta_{n+1}=\theta_{n}+\gamma_{n+1} \nabla\log
p_{\theta
}(y_{n}\mid y_{0\dvtx n-1})\mid_{\theta=\theta_{n}},\label
{equpdatethetapartialgradient}
\end{equation}
where the positive nonincreasing step-size sequence $ \{ \gamma
_{n} \} _{n\geq1}$
satisfies $\sum_{n}\gamma_{n}=\infty$ and $\sum_{n}\gamma
_{n}^{2}<\infty$
\cite{BMP90,legland1997}, for example, $\gamma
_{n}=n^{-\alpha}$
for $0.5<\alpha\leq1$. Upon receiving $y_{n}$, the parameter estimate
is updated in the direction of ascent of the conditional density of
this new observation. In other words, one recognizes in (\ref
{equpdatethetapartialgradient})
the update of the gradient ascent algorithm (\ref{eqbatchgradient}),
except that the partial (up to time $n$) likelihood is used. The
algorithm in the present form is, however, not suitable for on-line
implementation, because evaluating the gradient of $\log p_{\theta
}(y_{n}\mid y_{0\dvtx n-1})$
at the current parameter estimate requires computing the filter from
time $0$ to time $n$ using the current parameter value $\theta_{n}$.

An algorithm bypassing this problem has been proposed in the literature
for a finite state-space latent process in \cite{legland1997}. It
relies on the following update scheme:
%
\begin{equation}
\qquad \theta_{n+1}=\theta_{n}+\gamma_{n+1}\nabla\log
p_{\theta
_{0\dvtx n}}(y_{n}\mid y_{0\dvtx n-1}),\label{eqRML}
\end{equation}
where $\nabla\log p_{\theta_{0\dvtx n}}(y_{n}\mid y_{0\dvtx n-1})$ is defined as
%
\begin{eqnarray}\label{eqtimevaryingscore}
&& \nabla\log p_{\theta_{0\dvtx n}}(y_{n}\mid y_{0\dvtx n-1})
\nonumber\\[-8pt]\\[-8pt]\nonumber
&&\quad =\nabla\log p_{\theta
_{0\dvtx n}}(y_{0\dvtx n})-\nabla\log p_{\theta_{0\dvtx n-1}}(y_{0\dvtx n-1}),\hspace*{-20pt}
\end{eqnarray}
with the notation $\nabla\log p_{\theta_{0\dvtx n}}(y_{0\dvtx n})$ corresponding
to a ``time-varying'' score which is computed with a filter using the
parameter $\theta_{p}$ at time $p$. The update rule (\ref{eqRML})
can be thought of as an approximation to the update rule (\ref
{equpdatethetapartialgradient}).
If we use Fisher's identity to compute this ``time-varying'' score,
then we have for $1\leq p\leq n$,
%
\begin{eqnarray}\label{eqadditivetimevaryingscore}
s_{p}(x_{p-1\dvtx p}) &=& \nabla\log f_{\theta} (
x_{p}\mid x_{p-1} )\mid_{\theta=\theta
_{p}}
\nonumber\\[-8pt]\\[-8pt]\nonumber
&&{} + \nabla\log
g_{\theta} ( y_{p}\mid x_{p} )\mid_{\theta=\theta_{p}}.
\end{eqnarray}
The asymptotic properties of the recursion (\ref{eqRML}) (i.e., the
behavior of $\theta_{n}$ in the limit as $n$ goes to infinity)
has been studied in \cite{legland1997} for a finite state-space HMM.
It is shown that under regularity conditions this algorithm converges
toward a local maximum of the average log-likelihood $\ell
(\theta)$,
$\ell(\theta)$ being maximized at the ``true'' parameter
value under identifiability assumptions. Similar results hold for
the recursion (\ref{equpdatethetapartialgradient}).

\subsubsection{On-line Expectation--Maximization}

It is also possible to propose an on-line version of the EM algorithm.
This was originally proposed for finite state-space and linear Gaussian
models in \cite{elliott2000,ford1998}; see \cite{cappe11jcgs}
for a detailed presentation in the finite state-space case. Assume
that $p_{\theta}(x_{0\dvtx n},y_{0\dvtx n})$ is in the exponential family.
In the on-line implementation of EM, running averages of the sufficient
statistics $n^{-1}\mathcal{S}_{n}^{\theta}$ are computed \cite
{Cap09,elliott2000}. Let $\{\theta_{p}\}_{0\leq p\leq n}$ be the
sequence of parameter estimates of the on-line EM algorithm computed
sequentially based on $y_{0\dvtx n-1}$. When $y_{n}$ is received, we
compute
%
\begin{eqnarray}\label{eqsuffStatOnline}
\mathcal{S}_{\theta_{0\dvtx n}} &=& \gamma_{n+1} \int
s_{n} (x_{n-1\dvtx n} )\nonumber
\\
&&\hspace*{31pt}{}\cdot p_{\theta _{0\dvtx n}}(x_{n-1},x_{n}
\mid y_{0\dvtx n})\,dx_{n-1\dvtx n}
\nonumber\\[-8pt]\\[-8pt]\nonumber
&&{} + (1-\gamma_{n+1} )\sum_{k=0}^{n}
\Biggl(\prod_{i=k+2}^{n} (1-\gamma_{i} )
\Biggr)\gamma_{k+1}
\\
&&\hspace*{11pt}{}\cdot \int s_{k} (x_{k-1\dvtx k}
)p_{\theta_{0\dvtx k}}(x_{k-1\dvtx k}\mid y_{0\dvtx k})\,dx_{k-1\dvtx k},\nonumber
\end{eqnarray}
where $ \{ \gamma_{n} \} _{n\geq1}$ needs to satisfy $\sum_{n}\gamma
_{n}=\infty$
and $\sum_{n}\gamma_{n}^{2}<\infty$. Then the standard maximization
step (\ref{eqmaximiEM}) is used as in the batch version
%
\begin{equation}
\theta_{n+1}=\Lambda(\mathcal{S}_{\theta_{0\dvtx n}} ).
\end{equation}
The recursive calculation of $\mathcal{S}_{\theta_{0\dvtx n}}$ is achieved
by setting $V_{\theta_{0}}=0$, then computing
%
\begin{eqnarray}\label{equpdateStatonline}
V_{\theta_{0\dvtx n}} (x_{n} )
&=&\int\bigl\{ \gamma_{n+1} s_{n} (x_{n-1},x_{n} )\nonumber
\\
&&\hspace*{13pt}{}+ (1-\gamma_{n+1}
) V_{\theta_{0\dvtx n-1}} (x_{n-1} ) \bigr\}
\\
&&\hspace*{8pt}{} \cdot p_{\theta_{0\dvtx n}} (x_{n-1}\mid y_{0\dvtx n-1},x_{n} )\,dx_{n-1}\nonumber
\end{eqnarray}
and, finally,
%
\begin{equation}
\quad \mathcal{S}_{\theta_{0\dvtx n}}=\int V_{\theta_{0\dvtx n}} (x_{n}
)p_{\theta_{0\dvtx n}}(x_{n}\mid y_{0\dvtx n})\,dx_{n}.\label{equpdateStatonline2}
\end{equation}
Again, the subscript $\theta_{0\dvtx n}$ on $p_{\theta_{0\dvtx n}}(x_{0\dvtx n}\mid y_{0\dvtx n})$
indicates that the posterior density is being computed sequentially
using the parameter $\theta_{p}$ at time $p\leq n$. The filtering
density then is advanced from time $n-1$ to time $n$ by using
$f_{\theta_{n}}(x_{n}\mid x_{n-1})$,
$g_{\theta_{n}}(y_{n}\mid x_{n})$ and $p_{\theta_{n}}(y_{n}\mid y_{0\dvtx n})$
in the fraction of the r.h.s. of (\ref{eqjointupdate}). Whereas the
convergence of the EM algorithm toward a local maximum of the average
log-likelihood $\ell(\theta)$ has been \mbox{established} for
i.i.d. data \cite{cappe2009JRSSB}, its convergence for state-space
models remains an open problem despite empirical evidence it does
\cite{Cap09,cappe11jcgs,delmoralforward}. This
has motivated the \mbox{development} of modified versions of the on-line
EM algorithm for which convergence results are easier to establish
\cite{ADT05,lecorfffort2013}. However, the on-line EM presented
here usually performs empirically better \cite{lecorfffort2013-2}.

\subsubsection{Discussion of particle implementations}\label
{subDiscussion-of-particle-online}

Both the on-line gradient and EM procedures require approximating
terms (\ref{eqtimevaryingscore}) and (\ref{eqsuffStatOnline})
of the form (\ref{eqexpectationadditivefunctionals}), except that
the expectation is now w.r.t. the posterior density $p_{\theta
_{0\dvtx n}}(x_{0\dvtx n}\mid y_{0\dvtx n})$
which is updated using the parameter $\theta_{p}$ at time $p\leq n$.
In this on-line framework, only the path space, fixed-lag smoothing
and forward smoothing estimates are applicable; the fixed-lag approximation
is applicable but introduces a nonvanishing bias. For the on-line
EM algorithm, similarly to the batch case discussed in \mbox{Section}~\ref
{subDiscussion-of-particle-batch},
the benefits of using the forward smoothing estimate \cite{delmoralforward}
compared to the path space estimate \cite{Cap09} with $N^{2}$ particles
are rather limited, as experimentally demonstrated in Section~\ref
{subMaximum-likelihood-methods}.
However, for the on-line gradient ascent algorithm, the gradient term
$\nabla\log p_{\theta_{0\dvtx n}}(y_{n}|y_{0\dvtx n-1})$ in (\ref{eqRML})
is a difference between two score-like vectors (\ref{eqtimevaryingscore})
and the behavior of its particle estimates differs significantly from
its EM counterpart. Indeed, the variance of the particle path estimate
of $\nabla\log p_{\theta_{0\dvtx n}}(y_{n}|y_{0\dvtx n-1})$ increases linearly
with~$n$, yielding an unreliable gradient ascent procedure, whereas
the particle forward smoothing estimate has a variance uniformly bounded
in time under appropriate regularity assumptions and yields a stable
gradient ascent procedure \cite{pdmadsssfilterderivative}. Hence,
the use of a procedure of computational complexity $\mathcal{O}
(N^{2} )$
is clearly justified in this context. The very recent paper \cite
{westerbornOlsson2014}
reports that the computationally cheaper \mbox{estimate} (\ref
{eqOlssonMCapproximation})
appears to exhibit similar properties whenever $K\geq2$ and might
prove an attractive alternative.

\section{Bayesian Parameter Estimation}

\label{secBayesianestimation}

In the Bayesian setting, we assign a suitable prior density $p
(\theta)$
for $\theta$ and inference is based on the joint posterior density
$p ( x_{0\dvtx T},\theta\mid y_{0\dvtx T} )$ in the off-line
case or the sequence of posterior densities $ \{ p ( x_{0\dvtx n},\theta
\mid y_{0\dvtx n} ) \} _{n\geq0}$
in the on-line case.

\subsection{Off-Line Methods}

\subsubsection{Particle Markov chain Monte Carlo methods}\label{subPMCMC}

Using MCMC is a standard approach to approximate $p ( x_{0\dvtx T},\theta
\mid y_{0\dvtx T} )$. Unfortunately, designing
efficient MCMC sampling
algorithms for nonlinear \mbox{non-}Gaussian state-space models is a difficult
task:  one-variable-at-a-time Gibbs sampling typically mixes very poorly
for such models, whereas blocking strategies that have been proposed
in the literature are typically very model-dependent; see, for instance,
\cite{KimShephardChib}.

Particle MCMC are a class of MCMC techniques which rely on particle methods
to build efficient high-dimensional proposal distributions in a generic
manner~\cite{andrieu2010}. We limit ourselves here to the presentation
of the Particle Marginal Metropolis--Hastings (PMMH) sampler, which
is an approximation of an ideal MMH sampler for sampling from $p
( x_{0\dvtx T},\theta\mid y_{0\dvtx T} )$
which would utilize the following proposal density:
%
\begin{eqnarray}\label{eqproposalMMH}
&& q \bigl( \bigl(x_{0\dvtx T}^{\prime},\theta^{\prime} \bigr)\mid
(x_{0\dvtx T},\theta) \bigr)
\nonumber\\[-8pt]\\[-8pt]\nonumber
&&\quad =q \bigl( \theta^{\prime}\mid\theta
\bigr)p_{\theta^{\prime}} \bigl( x_{0\dvtx T}^{\prime}\mid y_{0\dvtx T}
\bigr),
\end{eqnarray}
where $q ( \theta^{\prime}\mid\theta)$ is
a proposal density to obtain a candidate $\theta^{\prime}$ when we
are at location $\theta$. The acceptance probability of this sampler
is
%
\begin{equation}
1\wedge\frac{p_{\theta^{\prime}} (y_{0\dvtx T} )p(\theta
^{\prime})q(\theta\mid\theta^{\prime})}{p_{\theta}
(y_{0\dvtx T} )p(\theta)q(\theta^{\prime}\mid\theta)}.\label{eqacceptprobaMH}
\end{equation}
Unfortunately, this ideal algorithm cannot be implemented, as we cannot
sample exactly from $p_{\theta^{\prime}} ( x_{0\dvtx T}\mid\break  y_{0\dvtx T} )$
and we cannot compute the likelihood terms $p_{\theta}
(y_{0\dvtx T} )$
and $p_{\theta^{\prime}} (y_{0\dvtx T} )$ appearing in the acceptance
probability.

The PMMH sampler is an approximation of this ideal MMH sampler
which relies on the particle approximations of these unknown terms.
Given $\theta$ and a particle approximation $\hat{p}_{\theta
} (y_{0\dvtx T} )$
of $p_{\theta} (y_{0\dvtx T} )$, we sample $\theta^{\prime
}\sim q ( \theta^{\prime}\mid\theta)$,
then run a particle filter to obtain approximations $\hat{p}_{\theta
^{\prime}} ( dx_{0\dvtx T}\mid y_{0\dvtx T} )$
and $\hat{p}_{\theta^{\prime}} (y_{0\dvtx T} )$ of
$p_{\theta^{\prime}} ( dx_{0\dvtx T}\mid y_{0\dvtx T} )$
and $p_{\theta^{\prime}} (y_{0\dvtx T} )$. We then sample
$X_{0\dvtx T}^{\prime}\sim\hat{p}_{\theta^{\prime}} ( dx_{0\dvtx T}\mid y_{0\dvtx T} )$,
that is, we choose randomly one of $N$ particles generated by the
particle filter, with probability $W_{T}^{i}$ for particle $i$,
and accept $ (\theta^{\prime},X_{0\dvtx T}^{\prime} )$ [and
$\hat{p}_{\theta^{\prime}} (y_{0\dvtx T} )$]
with probability
%
\begin{equation}
1\wedge\frac{\hat{p}_{\theta^{\prime}} (y_{0\dvtx T}
)p(\theta^{\prime})q(\theta\mid\theta^{\prime})}{\hat{p}_{\theta
} (y_{0\dvtx T} )p(\theta)q(\theta^{\prime}\mid\theta)}.\label
{eqacceptprobaPMMH}
\end{equation}
The acceptance probability (\ref{eqacceptprobaPMMH}) is a simple
approximation of the ``ideal'' acceptance probability
(\ref{eqacceptprobaMH}).

This algorithm was first proposed as a heuristic to sample from $p
( \theta\mid y_{0\dvtx T} )$
in \cite{fernandez2007}. Its remarkable feature established in \cite
{andrieu2010}
is that it does admit $p ( x_{0\dvtx T},\theta\mid
y_{0\dvtx T} )$
as invariant distribution whatever the number of particles $N$
used in the particle approximation \cite{andrieu2010}. However,
the choice of $N$ has an impact on the performance of the algorithm.
Using large values of $N$ usually results in PMMH averages with variances
lower than the corresponding averages using fewer samples, but the
computational cost of constructing $\hat{p}_{\theta}
(y_{0\dvtx T} )$
increases with $N$. A simplified analysis of this algorithm suggests
that $N$ should be selected such that the standard deviation of the
logarithm of the particle likelihood estimate should be around $0.9$
if the ideal MMH sampler was using the perfect proposal $q(\theta
^{\prime}|\theta)=p (\theta^{\prime}|y_{0\dvtx n} )$
\cite{pitt2011} and around $1.8$ if one uses an isotropic normal
random walk proposal for a target that is a
product of~$d$ i.i.d. components with  $d\rightarrow\infty$ \cite{sherlock2013}.
For general proposal and target densities, a recent theoretical analysis
and empirical results suggest that this standard deviation should
be selected around 1.2--1.3 \cite{doucet2012}. As the variance
of this estimate typically increases linearly with~$T$, this means
that the computational complexity is of order $\mathcal{O}(T^{2})$
by iteration.

A particle version of the Gibbs sampler is also available \cite{andrieu2010}
which mimicks the two-component Gibbs sampler sampling iteratively
from $p ( \theta\mid\break  x_{0\dvtx T},y_{0\dvtx T} )$ and
$p_{\theta} ( x_{0\dvtx T}\mid y_{0\dvtx T} )$. These
algorithms rely on a nonstandard version of the particle filter where
$N-1$ particles are generated conditional upon a ``fixed'' particle.
Recent improvements over this particle Gibbs sampler introduce mechanisms
to rejuvenate the fixed particle, using forward or backward sampling
procedures \mbox{\cite{whiteley2010backwarddiscussion,Lindsten2014,whiteley2010backward}}.
These methods perform empirically extremely
well, but, contrary to the PMMH, it is still unclear how one should
scale $N$ with $T$.

\subsection{On-Line Methods}
\label{subBayesianOn-line-Methods}

In this context, we are interested in approximating on-line the sequence
of posterior densities $ \{ p ( x_{0\dvtx n},\theta\mid  y_{0\dvtx n} ) \} _{n\geq0}$.
We emphasize that, contrary to the on-line ML parameter estimation
procedures, none of the methods presented in this section bypass the
particle degeneracy problem. This should come as no surprise. As discussed
in Section~\ref{secconvergenceparticle}, even for a \textit{fixed}
$\theta$, the particle estimate of $p_{\theta} (y_{0\dvtx n} )$
has a relative variance that increases linearly with $n$ under favorable
mixing assumptions. The methods in this section attempt to approximate
$p (\theta|y_{0\dvtx n} )\propto p_{\theta}(y_{0\dvtx n})p(\theta)$.
This is a harder problem, as it implicitly requires having to approximate
$p_{\theta^{i}} (y_{0\dvtx n} )$ for all the particles $ \{
\theta^{i} \} $
approximating $p (\theta|y_{0\dvtx n} )$.

\subsubsection{Augmenting the state with the parameter}

At first sight, it seems that estimating the sequence of posterior
densities $ \{ p ( x_{0\dvtx n},\theta\mid
y_{0\dvtx n} ) \} _{n\geq0}$
can be easily achieved using standard particle methods by merely
introducing the extended state $Z_{n}= (X_{n},\theta_{n} )$,
with initial density $p (\theta_{0} )\mu_{\theta
_{0}} (x_{0} )$
and transition density $f_{\theta_{n}} ( x_{n}\mid
x_{n-1} )\delta_{\theta_{n-1}} (\theta_{n} )$,
that is, $\theta_{n}=\theta_{n-1}$. However, this extended process $Z_{n}$
clearly does not possess any \emph{forgetting} property (as discussed
in Section~\ref{secfilteringandparticle}), so the algorithm is bound
to degenerate. Specifically, the parameter space is explored only
in the initial step of the algorithm. Then, each successive resampling
step reduces the diversity of the sample of $\theta$ values; after
a certain time $n$, the approximation $\hat{p} ( d\theta\mid y_{0\dvtx n} )$
contains a single unique value for $\theta$. This is clearly a poor
approach. Even in the much simpler case when there is no latent
variable $X_{0\dvtx n}$, it is shown in \cite{chopin2004}, Theorem 4,
that the asymptotic variance of the corresponding particle estimates
diverges at least at a polynomial rate, which grows with the dimension
of $\theta$.

A pragmatic approach that has proven useful in some applications is
to introduce artificial dynamics for the parameter $\theta$ \cite{Kita98},
%
\begin{equation}
\theta_{n+1}=\theta_{n}+\varepsilon_{n+1},
\end{equation}
where $ \{ \varepsilon_{n} \} _{n\geq0}$ is an artificial
dynamic noise with decreasing variance. Standard particle methods
can now be applied to approximate $ \{ p ( x_{0\dvtx n},\theta_{0\dvtx n}\mid\break 
y_{0\dvtx n} ) \} _{n\geq0}$.
A~related kernel density estimation method also appeared in \cite{Liu01},
which proposes to use a kernel density estimate $p (\theta
|y_{0\dvtx n} )$
from which one samples from. As before, the static parameter is transformed
to a slowly time-varying one, whose dynamics is related to the kernel
bandwidth. To mitigate the artificial variance \mbox{inflation}, a shrinkage
correction is introduced. An improved version of this method has been
recently proposed in~\cite{flury2010}.

It is difficult to quantify how much bias is introduced in the resulting
estimates by the introduction of this artificial dynamics. Additionally,
these methods require a significant amount of tuning, for example, choosing
the variance of the artificial dynamic noise or the kernel width.
However, they can perform satisfactorily in practice \cite{flury2010,Liu01}.

\subsubsection{Practical filtering}

The practical filtering approach proposed in \cite{Polson2002} relies
on the following fixed-lag approximation:
%
\begin{equation}
\qquad p ( x_{0\dvtx n-L},\theta\mid y_{0\dvtx n-1} )\approx p (
x_{0\dvtx n-L},\theta\mid y_{0\dvtx n} )
\end{equation}
for $L$ large enough; that is, observations coming after $n-1$ presumably
bring little information on $x_{0\dvtx n-L}$. To sample approximately
from $p ( \theta\mid y_{0\dvtx n} )$, one uses the
following iterative process:  at time $n$, several MCMC chains are
run in parallel to sample from
\begin{eqnarray*}
&& p \bigl( x_{n-L+1\dvtx n},\theta\mid y_{0\dvtx n},X_{0\dvtx n-L}^{i}
\bigr)
\\
&&\quad =p \bigl( x_{n-L+1\dvtx n},\theta\mid y_{n-L+1\dvtx n},X_{n-L}^{i}
\bigr),
\end{eqnarray*}
where the $X_{n-L}^{i}$ have been obtained at the previous iteration
and are such that (approximately) $X_{n-L}^{i}\sim
p(x_{n-L}|y_{0\dvtx n-1})\approx p(x_{n-L}|y_{0\dvtx n})$.
Then one collects the first component $X_{n-L+1}^{i}$ of the simulated
sample $X_{n-L+1\dvtx n}^{i}$, increments the time index and runs several
new MCMC chains in parallel to sample from $p ( x_{n-L+2\dvtx n+1},\theta
\mid y_{n-L+2\dvtx n+1},X_{n-L+1}^{i} )$
and so on. The algorithm is started at time $L-1$, with MCMC chains
that target $p(x_{0\dvtx L-1}|y_{0\dvtx L-1})$. Like all methods based on fixed-lag
approximation, the choice of the lag $L$ is difficult and this introduces
a nonvanishing bias which is difficult to quantify. However, the
method performs well on the examples presented in \cite{Polson2002}.

\subsubsection{Using MCMC steps within particle methods}\label
{subUsing-MCMCsteps-within-SMC}

To avoid the introduction of an artificial dynamic model or of a fixed-lag
approximation, an approach originally proposed independently in \cite{Fea02}
and \cite{Gilks01} consists of adding MCMC steps to re-introduce
``diversity'' among the particles. Assuming we use
an auxiliary particle filter to\vspace*{1pt} approximate $ \{ p ( x_{0\dvtx n},\theta
\mid y_{0\dvtx n} ) \} _{n\geq0}$,
then the particles $ \{ X_{0\dvtx n}^{i},\theta_{n}^{i} \} $
obtained after the sampling step at time $n$ are approximately distributed
according to
\begin{eqnarray*}
&& \tilde{p} ( x_{0\dvtx n},\theta\mid y_{0\dvtx n} )
\\
&&\quad \propto p (
x_{0\dvtx n-1},\theta\mid y_{0\dvtx n-1} )q_{\theta}
(x_{n},y_{n}|x_{n-1} ).
\end{eqnarray*}
We have $\tilde{p} ( x_{0\dvtx n},\theta\mid
y_{0\dvtx n} )=p ( x_{0\dvtx n},\theta\mid
y_{0\dvtx n} )$
if $q_{\theta} (x_{n}|\break y_{n},x_{n-1} )=p_{\theta}
(x_{n}\hspace*{-0.5pt}| y_{n},x_{n-1} )$
and $q_{\theta} (y_{n}|x_{n-1} )=p_{\theta}
(y_{n}\hspace*{-0.5pt}|\break  x_{n-1} )$.
To add diversity in this population of particles, we introduce an
MCMC kernel $K_{n} (d (x_{0\dvtx n}^{\prime},\theta^{\prime
} )|\break  (x_{0\dvtx n}, \theta) )$
with invariant density $\tilde{p} ( x_{0\dvtx n},\theta
\mid y_{0\dvtx n} )$
and replace, at the end of each iteration, the set of resampled particles,
$(\overline{X}_{0\dvtx n}^{i},\bar{\theta}_{n}^{i})$ with $N$ ``mutated''
particles $ (\widetilde{X}_{0\dvtx n}^{i},\tilde{\theta
}_{n}^{i} )$
simulated from, for $i=1,\ldots,N$,
\[
\bigl(\widetilde{X}_{0\dvtx n}^{i},\tilde{\theta}_{n}^{i}
\bigr)\sim K_{n} \bigl(d(x_{0\dvtx n},\theta)| \bigl(
\overline{X}_{0\dvtx n}^{i},\bar{\theta}_{n}^{i}
\bigr) \bigr).
\]
If we use the SISR algorithm, then we can alternatively use an MCMC
step of invariant density $p ( x_{0\dvtx n},\theta\mid
y_{0\dvtx n} )$
after the resampling step at time $n$.

Contrary to standard applications of MCMC, the kernel does not have
to be ergodic. Ensuring ergodicity would indeed require one to sample
an increasing number of variables as $n$ increases---this algorithm
would have an increasing cost per iteration, which would prevents
its use in on-line scenarios, but it can be an interesting alternative
to standard MCMC and was suggested in \cite{leedominic2002}. In
practice, one therefore sets $\widetilde{X}_{0\dvtx n-L}^{i}=X_{0\dvtx n-L}^{i}$
and only samples $\theta^{i}$ and $\widetilde{X}_{n-L+1\dvtx n}^{i}$,
where $L$ is a small integer; often $L=0$ (only $\theta$ is updated).
Note that the memory requirements for this method do not increase
over time if $\tilde{p}_{\theta} (x_{0\dvtx n},y_{0\dvtx n} )$
is in the exponential family and thus can be summarized by a set of
fixed-dimensional sufficient statistics $s^{n}(x_{0\dvtx n},y_{0\dvtx n})$.
This type of method was first used to perform on-line Bayesian
parameter estimation in a context where $\tilde{p}_{\theta}
(x_{0\dvtx n},y_{0\dvtx n} )$
is in the exponential family \cite{Gilks01,Fea02}. Similar
strategies were adopted in \cite{Andrieu99} and \cite{Sto02}. In
the particular scenario where $q_{\theta}
(x_{n}|y_{n},x_{n-1} )=p_{\theta}
(x_{n}|y_{n},x_{n-1} )$
and $q_{\theta} (y_{n}|x_{n-1} )=p_{\theta}
(y_{n}|x_{n-1} )$,
this method was mentioned in \cite{Andrieu99,vercauteren2005}
and is discussed at length in \cite{Lopes2010} who named it particle
learning. Extensions of this strategy to parameter estimation in conditionally
linear Gaussian models, where a part of the state is integrated out
using Kalman techniques \cite{Chen00,Douc00}, is proposed
in \cite{Car10}.

As opposed to the methods relying on kernel or artificial dynamics,
these MCMC-based approaches have the advantage of adding diversity
to the particles approximating $p (\theta|y_{0\dvtx n} )$ without
perturbing the target distribution. Unfortunately, these algorithms
rely implicitly on the particle approximation of the density $p
(x_{0\dvtx n}|y_{0\dvtx n} )$
even if algorithmically it is only necessary to store some fixed-dimensional
sufficient statistics $ \{ s^{n}(X_{0\dvtx n}^{i},y_{0\dvtx n}) \} $.
Hence, in this respect they suffer from the degeneracy problem. This
was noticed as early as in \cite{Andrieu99}; see also the word of
caution in the conclusion of \cite{ADT05,Fea02} and \cite
{PLValenciaDiscussion}.
The practical implications are that one observes empirically that
the resulting Monte Carlo estimates can display quite a lot of variability
over multiple runs as demonstrated in Section~\ref
{subbayesianmethodsexperiments}.
This should not come as a surprise, as the \mbox{sequence} of posterior distributions
does not have exponential forgetting properties, hence, there is an
accumulation of Monte Carlo errors over time.

\subsubsection{The SMC$^{2}$ algorithm}

The SMC$^{2}$ algorithm introduced simultaneously in \cite{smc2}
and \cite{fulop2013} may be considered as the particle equivalent
of Particle MCMC. It mimics an ``ideal'' particle algorithm proposed
in \cite{chopin2002} approximating sequentially $ \{ p
( \theta\mid y_{0\dvtx n} ) \} _{n\geq0}$
where $N_{\theta}$ particles (in the $\theta$-space) are used to
explore these distributions. The $N_{\theta}$ particles at time $n$
are reweighted according to $p_{\theta}(y_{0\dvtx n+1})/p_{\theta}(y_{0\dvtx n})$
at time $n+1$. As these likelihood terms are unknown, we substitute
to them $\hat{p}_{\theta}(y_{0\dvtx n+1})/\hat{p}_{\theta}(y_{0\dvtx n})$ where
$\hat{p}_{\theta}(y_{0\dvtx n})$ is a particle approximation of the partial
likelihood $p_{\theta}(y_{0\dvtx n})$, obtained by a running a particle
filter of $N_{x}$ particles in the $x$-dimension, up to time $n$,
for each of the $N_{\theta}$ $\theta$-particles. When particle degeneracy
(in the $\theta$-dimension) reaches a certain threshold, $\theta$-particles
are refreshed through the succession of a resampling step, and an
MCMC step, which in these particular settings takes the form of a
PMCMC update. The cost per iteration of this algorithm is not constant
and, additionally, it is advised to increase $N_{x}$ with $n$ for
the relative variance of $\hat{p}_{\theta}(y_{0\dvtx n})$ not to increase,
therefore, it cannot be used in truly on-line scenarios. Yet there
are practical situations where it may be useful to approximate jointly
all the posteriors $p(\theta|y_{1\dvtx n})$, for $1\leq n\leq T$, for
instance, to assess the predictive power of the model.

\section{Experimental Results}\label{secexperimentalresults}

We focus on illustrating numerically a few algorithms and the impact
of the degeneracy problem on parameter inference. This last point
is motivated by the fact that particle degeneracy seems to have been
overlooked by many practitioners. In this way numerical results may
provide valuable insights.

We will consider the following simple scalar linear Gaussian state
space model:
%
\begin{equation}
X_{n}=\rho X_{n-1}+\tau W_{n},\quad
Y_{n}=X_{n}+\sigma V_{n},\label{eqdlm1}
\end{equation}
where $V_{n}, W_{n}$ are independent zero-mean and unit-variance
Gaussians and $\rho\in[-1,1]$. The main reason for choosing this
model is that Kalman  recursions can be implemented to provide
the exact values of the summary statistics $\mathcal{S}_{n}^{\theta}$
used for ML estimation through the EM algorithm and to compute the
exact likelihood $p_{\theta} (y_{0\dvtx n} )$. Hence, using a
fine discretization of the low-dimensional parameter space, we can
compute a very good approximation of the true posterior density $p
( \theta\mid y_{0\dvtx n} )$.
In this model it is straightforward to present numerical evidence
of some effects of degeneracy for parameter estimation and to show how
it can be overcome by choosing an appropriate particle method.

%
\begin{figure*}

\includegraphics{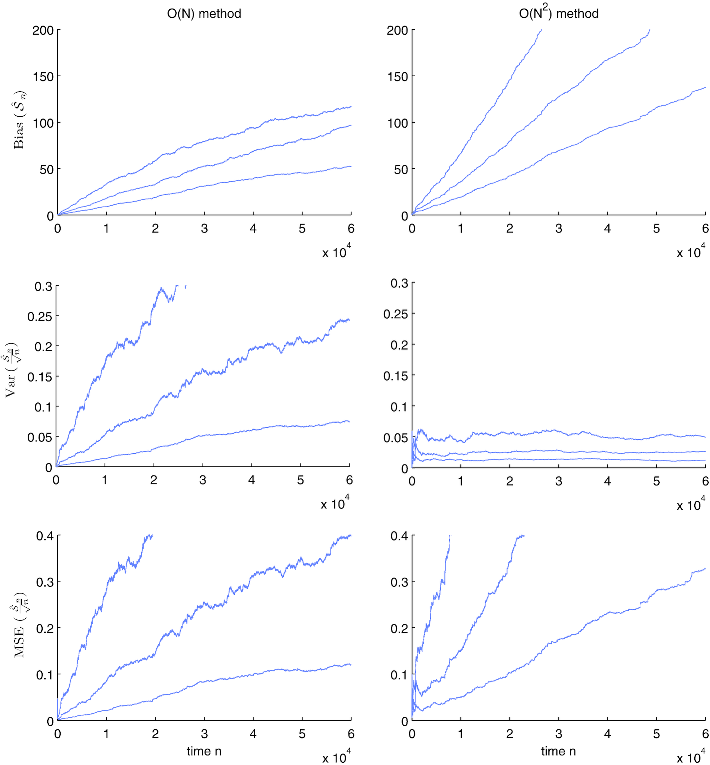}

\caption{Estimating smoothed additive functionals: empirical bias of
the estimate
of $\mathcal{S}_{n}^{\theta}$ (top panel), empirical variance (middle
panel) and MSE (bottom panel) for the estimate of $\mathcal
{S}_{n}^{\theta}/\sqrt{n}$.
Left column: $\mathcal{O}(N)$ method using $N^{2}={}$2500, 10,000, 40,000
particles. Right column: $\mathcal{O}(N^{2})$ method using $N={}$50, 100, 200
particles. In every subplot, the top line corresponds to using $N=50$,
the middle for $N=100$ and the lower for $N=200$.}\label{figsmoothing}
\end{figure*}

\subsection{Maximum Likelihood Methods}\label{subMaximum-likelihood-methods}

As\vspace*{1pt} ML methods require approximating smoothed additive functionals
$\mathcal{S}_{n}^{\theta}$ of the form (\ref
{eqexpectationadditivefunctionals}),
we begin by investigating the empirical bias, variance and MSE of
two standard particle estimates of $\mathcal{S}_{n}^{\theta}$, where
we set $s_{k}(x_{k-1},x_{k})=x_{k-1}x_{k}$ for the model described
in~(\ref{eqdlm1}). The first estimate relies on the path space method
with computational cost $\mathcal{O}(N)$ per time, which uses
$\hat{p}_{\theta} ( dx_{0\dvtx n}\mid y_{0\dvtx n} )$
in (\ref{eqSMCfullPosterior}) to approximate $\mathcal{S}_{n}^{\theta}$
as $\widehat{\mathcal{S}}_{n}^{\theta}$; see \cite{cappe2005}, Section~8.3,
for more details. The second estimate relies on the forward implementation
of FFBSm presented in Section~\ref{subForward-only-smoothing} using
(\ref{eqadditivesmoothfunctionalsasfunctionofT})--(\ref
{eqSMCapproxadditivefunctionals-1});
see \cite{delmoralforward}. Recall that this procedure has a computational
cost that is $\mathcal{O}(N^{2})$ per time for $N$ particles and
provides the same estimates as the standard forward--backward implementation
of FFBSm. For the sake of brevity, we will not consider the remaining
smoothing methods of Section~\ref{secsmoothingandparticle}; for
the fixed-lag and the exponentially weighted approximations we refer
the reader to \cite{olsson2008}, respectively, \cite{nemeth2013} for
numerical experiments.

We use a simulated data set of size $6\times10^{4}$ obtained using
$\theta^{\ast}=(\rho^{\ast},\tau^{2^{*}},\sigma^{2^{*}})=(0.8,0.1,1)$
and then generate 300 independent replications of each method in order
to compute the empirical bias and variance of $\widehat
{\mathcal{S}}_{n}^{\theta^{*}}$
when $\theta$ is fixed to $\theta^{\ast}$. In order to make a comparison
that takes into account the computational cost, we use $N^{2}$ particles
for the $\mathcal{O}(N)$ method and $N$ for the $\mathcal{O}(N^{2})$
one. We look separately at the behavior of the bias of $\widehat
{\mathcal
{S}}_{n}^{\theta}$
and the variance\vspace*{1pt} and MSE of the rescaled estimates $\widehat{\mathcal
{S}}_{n}^{\theta}/\sqrt{n}$.
The results are presented in Figure~\ref{figsmoothing} for $N={}$50, 100, 200.

For both methods the bias grows linearly with time, this growth being
higher for the $\mathcal{O}(N^{2})$ method. For the variance of
$\widehat
{\mathcal{S}}_{n}^{\theta}/\sqrt{n}$,
we observe a linear growth with time for the $\mathcal{O}(N)$ method
with $N^{2}$ particles, whereas this variance appears roughly constant
for the $\mathcal{O}(N^{2})$ method. Finally, the MSE of $\widehat
{\mathcal{S}}_{n}^{\theta}/\sqrt{n}$
grows for both methods linearly as expected. In this particular scenario,
the constants of proportionality are such that the MSE is lower for
the $\mathcal{O}(N)$ method than for the $\mathcal{O}(N^{2})$ method.
In general, we can expect that the $\mathcal{O}(N)$ method be superior
in terms of the bias and the $\mathcal{O}(N^{2})$ method superior
in terms of the variance. These results are in agreement with the
theoretical results in the literature \cite{delmoral2009,delmoralforward,douc09}, but additionally show that the lower bound on
the variance
growth of $\widehat{\mathcal{S}}_{n}^{\theta}$ for the $\mathcal{O}(N)$
method of \cite{poyadjis2009} appears sharp.

%
\begin{figure*}[t]

\includegraphics{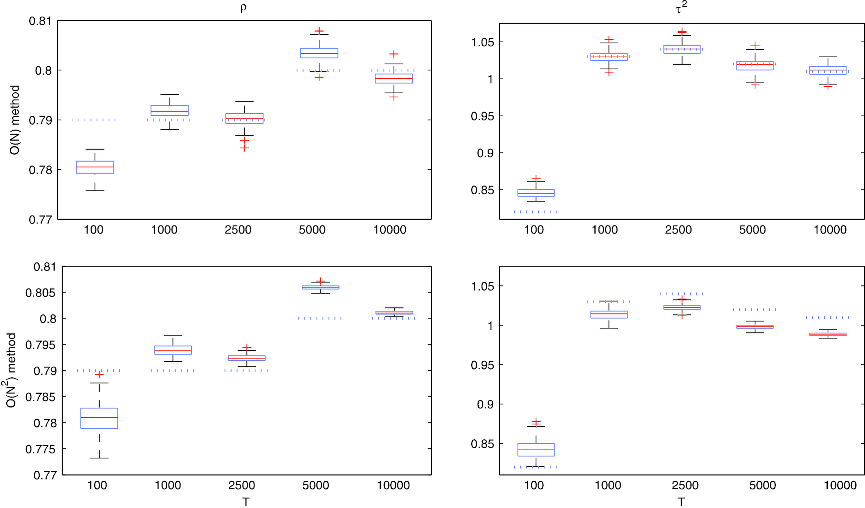}

\caption{Off-line EM:  boxplots of $\hat{\theta}_{n}$ for various $T$ using
25 iterations of off-line EM and 150 realizations of the algorithms.
Top panels: $\mathcal{O}(N)$ method using $N=150^{2}$ particles.
Bottom panels: $\mathcal{O}(N^{2})$ with $N=150$. The dotted horizontal
lines are the ML estimate for each time $T$ obtained using Kalman
filtering on a grid.}\vspace*{6pt}\label{figofflineEM}
\end{figure*}

%
\begin{figure*}[b]\vspace*{6pt}

\includegraphics{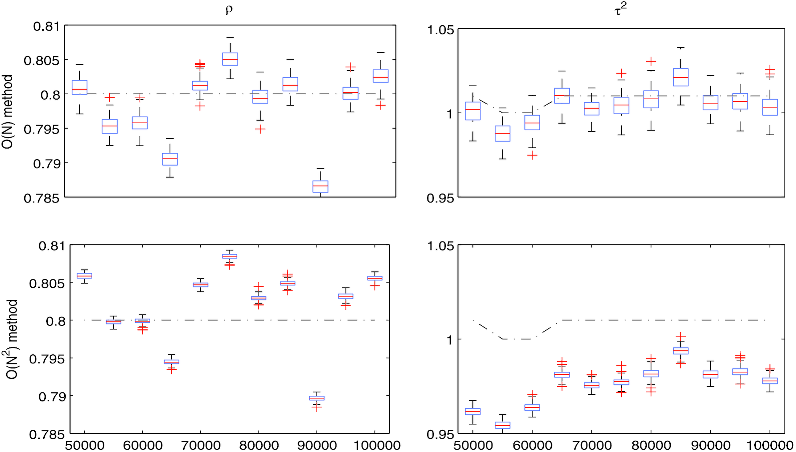}

\caption{On-line EM: boxplots of $\hat{\theta}_{n}$ for $n\geq
5\times10^{4}$
using 150 realizations of the algorithms. We also plot the ML estimate
at time $n$ obtained using Kalman filtering on a grid (black).}
\label{figonlineEM}
\end{figure*}

We proceed to see how the bias and variance of the estimates of
$\mathcal{S}_{n}^{\theta}$
affect the ML estimates, when the former are used within both an off-line
and an on-line EM algorithm; see Figures~\ref{figofflineEM} and
\ref{figonlineEM}, respectively. For the model in (\ref{eqdlm1})
the E-step corresponds to computing $\mathcal{S}_{n}^{\theta}$ where
$s_{k}(x_{k-1},x_{k})=
((y_{k}-x_{k})^{2},x_{k-1}^{2},x_{k-1}x_{k},x_{k}^{2} )$
and the M-step update function is given by
\[
\Lambda(z_{1},z_{2},z_{3},z_{4})=
\biggl(\frac
{z_{3}}{z_{4}},z_{4}-\frac{z_{3}^{2}}{z_{2}},z_{1}
\biggr).
\]
We compare the estimates of $\theta^{\ast}$ when the E-step is computed
using the $\mathcal{O}(N)$ and the $\mathcal{O}(N^{2})$ methods
described in the previous section with $150^{2}$ and $150$ particles,
respectively. A simulated data set for $\theta^{\ast}=(\rho^{\ast
},\tau^{*},\sigma^{*})=(0.8,1,0.2)$
will be used. In every case we will initialize the algorithm using
$\theta_{0}=(0.1,0.1,0.2)$ and assume $\sigma^{*}$ is known. In
Figures~\ref{figofflineEM} and \ref{figonlineEM} we present the
results obtained using 150 independent replications of the algorithm.
For the off-line EM, we use $25$ iterations for $T={}$100, 1000, 2500, 5000, 10,000.
For the on-line EM, we use $T=10^{5}$ with the step size set as $\gamma
_{n}=n{}^{-0.8}$
and for the first $50$ iterations no M-step update is performed.
This ``freezing'' phase is required to allow for a reasonable estimation
of the summary statistic; see \cite{Cap09,cappe11jcgs} for
more details. Note that in Figure~\ref{figonlineEM} we plot only
the results after the algorithm has converged, that is, for $n\geq5
\times10^{4}$.
In each case, both the $\mathcal{O}(N)$ and the $\mathcal{O}(N^{2})$
methods yield fairly accurate results given the low number of particles
used. However, we note, as observed previously in the literature, that
the on-line EM as well as the on-line gradient ascent method requires
a substantial number of observations, that is, over 10,000, before achieving
convergence \cite{Cap09,cappe11jcgs,delmoralforward,poyadjis2009}. For
smaller data sets, these algorithms can also
be used by going through the data, say, $K$ times. Typically, this method
is cheaper than iterating (\ref{eqbatchgradient}) or (\ref
{eqQfunction})--(\ref{eqmaximiEM})
$K$~times the off-line algorithms and can yield comparable parameter
estimates \cite{yildirim2013}. Experimentally, the properties of
the estimates of $\mathcal{S}_{n}^{\theta}$ discussed earlier appear
to translate into properties of the resulting parameter estimates:
the $\mathcal{O}(N)$ method provides estimates with less bias but
more variance than the $\mathcal{O}(N^{2})$ method.

For more numerical examples regarding the remaining methods discussed
in Section~\ref{secMLestimation}, we refer the reader to \cite
{Ionides09,ionidesbretoking2006} for iterated filtering, to \cite
{delmoralforward,delmoral2009,poyadjis2009} for comparisons of the
$\mathcal{O}(N)$ and $\mathcal{O}(N^{2})$ methods for EM and gradient
ascent, to \cite{Cap09} for the $\mathcal{O}(N)$ on-line EM, to
\cite{Pitt02} and \cite{Lee08}, Chapter~10, for smooth likelihood function
methods and to \cite{cappe2005}, Chapters~10--11, for a detailed exposition
of off-line EM methods.

\subsection{Bayesian Methods}\label{subbayesianmethodsexperiments}

We still consider the model in (\ref{eqdlm1}), but simplify it further
by fixing either $\rho$ or $\tau$. This is done in order to keep
the computations of the benchmarks that use Kalman computations on
a grid relatively inexpensive. For those parameters that are not fixed,
we shall use the following independent priors: a uniform on $[-1,1]$
for~$\rho$, and inverse gamma for $\tau^{2},\sigma^{2}$ with the
shape and scale parameter pair being $(a,b)$ and $(c,d)$, respectively,
with $a=b=c=d=1$. In all the subsequent examples, we will initialize
the algorithms by sampling $\theta$ from the prior.

%
\begin{figure*}

\includegraphics{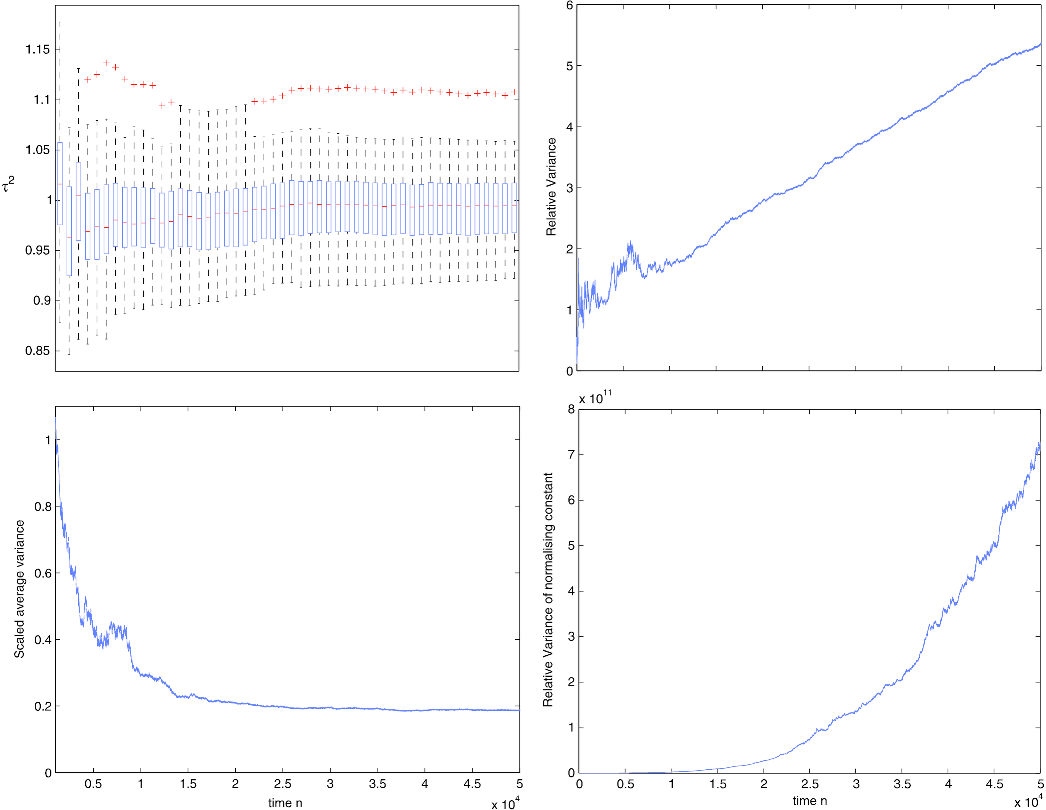}

\caption{Top left: box plots for estimates of posterior mean of $\tau^{2}$
at $n={}$1000, 2000$,\ldots,$50,000. Top right: relative variance, that is,
empirical variance (over independent runs) for the estimator of the
mean of $p(\tau^{2}|y_{0\dvtx n})$ using particle method with MCMC steps normalized
with the true posterior variance computed using Kalman filtering on
a grid. Bottom left: average (over independent runs) of the estimated
variance of $p(\tau^{2}|y_{0\dvtx n})$ using particle method with MCMC
normalized with the true posterior variance. Bottom right: relative variance
of the $ \{ \hat{p} (y_{0\dvtx n} ) \} _{n\geq0}$; All
plots are computed using $N=5000$ and over 100 different independent
runs.}\vspace*{6pt}
\label{FlorelVar}
\end{figure*}

We proceed to examine the particle algorithms with MCMC moves that
we described in Section~\ref{subUsing-MCMCsteps-within-SMC}.
We focus on an efficient implementation of this idea discussed in
\cite{Lopes2010} which can be put in practice for the simple model
under consideration. We investigate the effect of the degeneracy problem
in this context. The \mbox{numerical} results obtained in this section have
been produced in Matlab (code available from the first author) and
double-checked using the R program available on the personal web page
of the first author of~\mbox{\cite{Lopes2010,Lopezforecast}}.

We first focus on the estimate of the posterior of $\theta=(\tau
^{2},\sigma^{2})$
given a long sequence of simulated observations with $\tau=\sigma=1$.
In this scenario, $p_{\theta}(x_{0\dvtx n},y_{0\dvtx n})$ admits the following
two-dimensional sufficient statistics, $s^{n}(x_{0\dvtx n},y_{0\dvtx n})=
(\sum_{k=1}^{n} (x_{k}-x_{k-1} )^{2},\sum_{k=0}^{n}
(y_{k}-x_{k} )^{2} )$,
and $\theta$ can be updated using Gibbs steps. We use $T=5\times10^{4}$
and $N=5000$. We ran the algorithm over 100 independent runs over
the same data set. We present the results only for $\tau^{2}$ and
omit the ones for $\sigma^{2}$, as these were very similar. The top
left panel of Figure~\ref{FlorelVar} shows the box plots for the
estimates of the posterior mean, and the top right panel shows how
the corresponding relative variance of the estimator for the posterior
mean evolves with time. Here the relative variance is defined as the
ratio of the empirical variance (over different independent runs)
of the posterior mean estimates at time $n$ over the true posterior
variance at time $n$, which in this case is approximated using a
Kalman filter on a fine grid. This quantity exhibits a steep increasing
trend when $n\geq{}$15,000 and confirms the aforementioned variability
of the estimates of the posterior mean. In the bottom left panel of
Figure~\ref{FlorelVar} we plot the average (over different runs)
of the estimators of the variance of $p(\tau^2|y_{0\dvtx n})$. This average
variance is also scaled/normalized by the actual posterior variance.
The latter is again computed using Kalman filtering on a grid. This
ratio between the average estimated variance of the posterior over
the true one decreases with time $n$ and it shows that the supports
of the approximate posterior densities provided by this method cover,
on average, only a small portion of the support of the true posterior.
These experiments confirm that in this example the particle method
with MCMC steps fails to adequately explore the space of~$\theta$.
Although the box plots provide some false sense of security, the relative
and scaled average variance clearly indicate that any posterior estimates
obtained from a single run of particle method with MCMC steps should
be used with caution. Furthermore, in the bottom right panel of
Figure~\ref{FlorelVar} we also investigate experimentally the empirical
relative variance of the marginal likelihood estimates $ \{ \hat{p}
(y_{0\dvtx n} ) \} _{n\geq0}$.
This relative variance appears to increase quadratically with $n$ for the
particle method with MCMC moves instead of linearly as it does for
state-space models with good mixing properties. This suggests that
one should increase the number of particles quadratically with
the time index to obtain an estimate of the marginal likelihood whose
relative variance remains uniformly bounded with respect to the time
index. Although we attribute this quadratic relative variance growth to the
degeneracy problem, the estimate $\hat{p} (y_{0\dvtx n} )$ is
not the particle approximation of a smoothed additive functional,
thus there is not yet any theoretical convergence result explaining
rigorously this phenomenon.

%
\begin{figure*}

\includegraphics{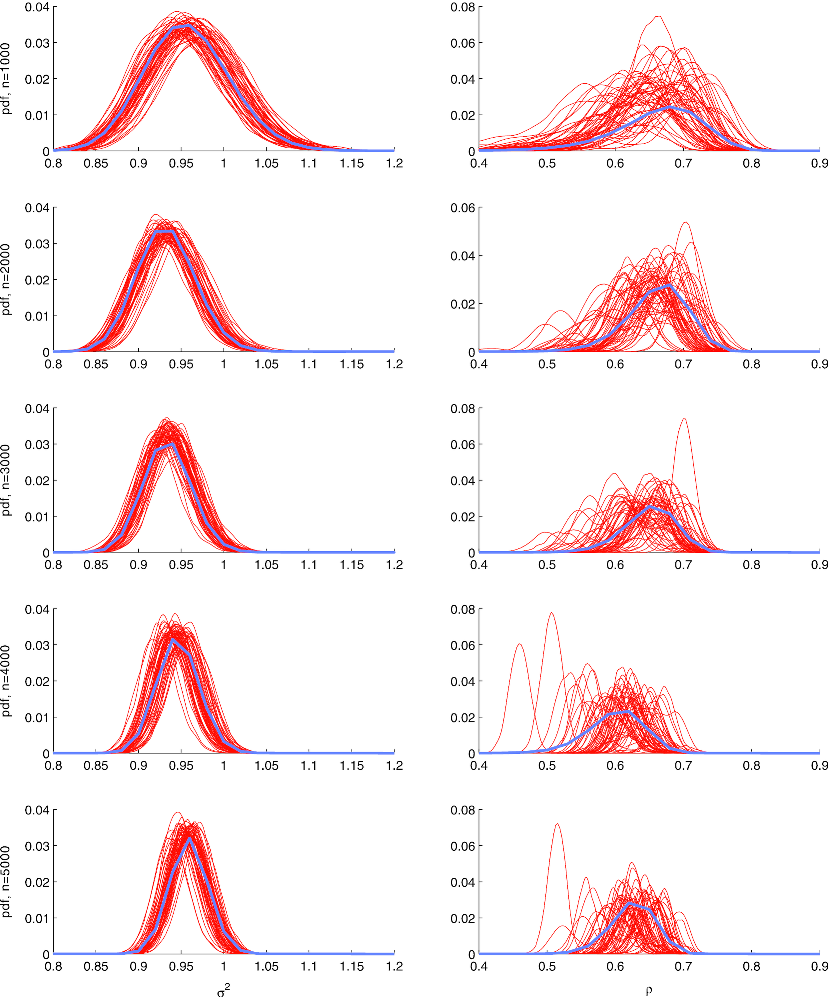}

\caption{Particle method with MCMC steps, $\theta=(\rho,\sigma
^{2})$; estimated
marginal posterior densities for $n=10^{3},2\times10^{3},\dots,5\times10^{3}$
over 50 runs (red) versus ground truth (blue).}\vspace*{-3pt}
\label{Flodlmdriftsnapshot}
\end{figure*}

One might argue that these particle methods with MCMC moves are meant
to be used with larger $N$ and/or shorter data sets $T$. We shall
consider this time a slightly different example where $\tau=0.1$
is known and we are interested in estimating the posterior of $\theta
=(\rho,\sigma^{2})$
given a sequence of observations obtained using $\rho=0.5$ and $\sigma=1$.
In that case, the sufficient statistics are
$s^{n}(x_{0\dvtx n},y_{0\dvtx n})= (\sum_{k=1}^{n}x_{k-1}x_{k},\sum
_{k=0}^{n-1}x_{k-1}^{2},\sum_{k=0}^{n} (y_{k}-x_{k}
)^{2} )$,
and the parameters can be rejuvenated through a single Gibbs update.
In addition, we let $T=5000$ and use $N=10^{4}$ particles. In
Figure~\ref{Flodlmdriftsnapshot} we display the estimated marginal posteriors
$p ( \rho\mid y_{0\dvtx n} )$ and $p ( \sigma^{2}\mid y_{0\dvtx n} )$
obtained from\ $50$ independent replications of the particle method.
On this simple problem, the estimated posteriors seem consistently
rather inaccurate for $\rho$, whereas they perform better for $\sigma^{2}$
but with some nonnegligible variability over runs, which increases
as $T$ increases. Similar observations have been reported in \cite
{PLValenciaDiscussion}
and remain unexplained: for some parameters this methodology appears
to provide reasonable results despite the degeneracy problem and for
others it provides very unreliable results.

%
\begin{figure*}[p]

\includegraphics{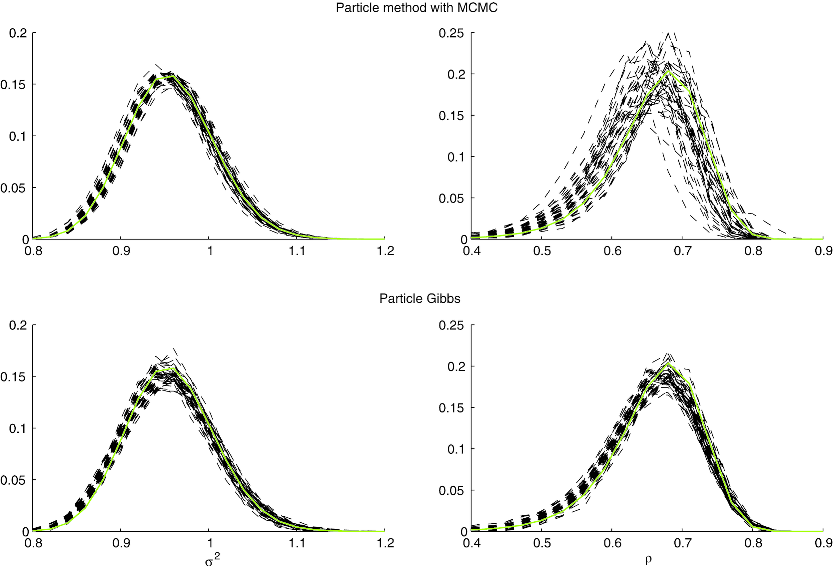}

\caption{Estimated marginal posterior densities for $\theta=(\rho,\sigma^{2})$
with $T=10^{3}$ over 50 runs (black-dashed) versus ground truth (green).
Top: particle method with MCMC steps, $N=7.5\times10^{4}$. Bottom: particle
Gibbs with $3000$ iterations and $N=50$.}\label{Flocomparelow}
\end{figure*}

%
\begin{figure*}[p]

\includegraphics{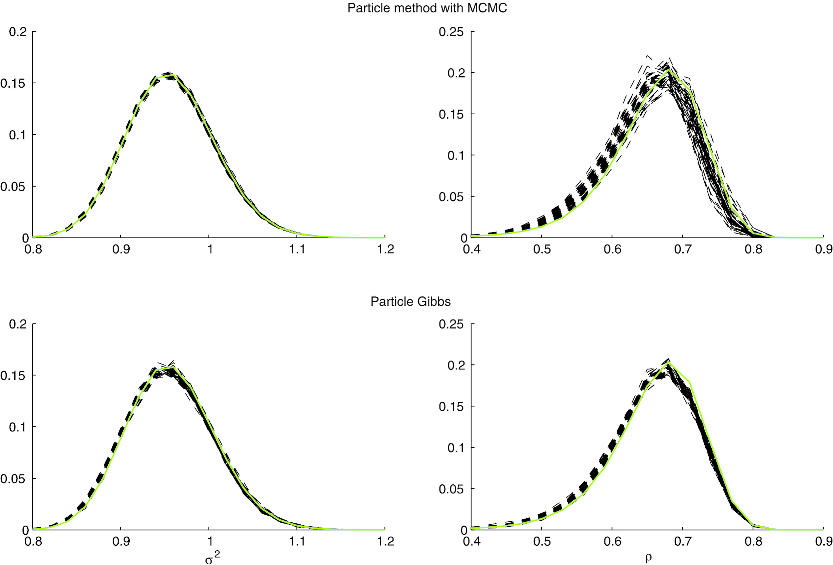}

\caption{Estimated marginal posterior densities for $\theta=(\rho,\sigma^{2})$
with $T=10^{3}$ over 50 runs (black-dashed) versus ground truth (green).
Top: particle method with MCMC steps, $N=6\times10^{5}$. Bottom: particle
Gibbs with 24,000 iterations and $N=50$.}\label{Flocomparehigh}
\end{figure*}

We investigate further the performance of this meth\-od in this simple
example by considering the same \mbox{example} for $T=1000$, but now consider
two larger numbers of particles, $N=7.5\times10^{4}$ and $N=6\times10^{5}$,
over 50 different runs. Additionally, we compare the resulting\vadjust{\goodbreak} estimates
with estimates provided by the particle Gibbs sampler of \cite{Lindsten2014}
using the same computational cost, that is, $N=50$ particles with
3000 and 24,000 iterations, respectively. The results are displayed
in Figures~\ref{Flocomparelow} and \ref{Flocomparehigh}. As expected,
we improve the performance of the particle with MCMC moves when $N$
increases for a fixed time horizon~$T$. For a fixed computational
complexity, the particle Gibbs sampler estimates appear to display
less variability. For a higher-dimensional parameter $\theta$ and/or
very vague priors, this comparison would be more favorable to the
particle Gibbs sampler as illustrated in \cite{andrieu2010}, pages~\mbox{336--338}.

\section{Conclusion}\label{secconclusion}

Most particle methods proposed originally in the literature to perform
inference about static parameters in general state-space models were
computationally inefficient as they suffered from the degeneracy problem.
Several approaches have been proposed to deal with this problem by
either adding an artificial dynamic on the static parameter \cite
{flury2009,Kita98,Liu01} or introducing a fixed-lag approximation
\cite{kitagawa2001,olsson2008,Polson2002}. These
methods can work very well in practice, but it remains unfortunately
difficult/impossible to quantify the bias introduced in most realistic
applications. Various asymptotically bias-free methods with good statistical
\mbox{properties} and a reasonable computational cost have recently appeared
in the literature.

To perform batch ML estimation, the forward filter backward sampler/smoother
and generalized two-filter procedures are recommended whenever the
$\mathcal{O}(N^{2}T)$ computational complexity per iteration of their
direct implementations can be lowered to $\mathcal{O}(NT)$ using,
for example, the methods described in \mbox{\cite{briers2005,douc09,fearnhead2008,Klass2005}}. Otherwise, besides a lowering
of memory requirements, not much can be gained from these techniques
compared to simply using a standard particle filter with $N^{2}$
particles. In an on-line ML context, the situation is markedly different.
Whereas for the on-line EM algorithm, the forward smoothing approach
in \mbox{\cite{delmoralforward,poyadjis2009}} of complexity
$\mathcal{O}(N^{2})$
per time step will be similarly of limited interest compared to a
standard particle filter using $N^{2}$ particles; it is crucial to
use this approach when performing on-line gradient ascent as demonstrated
empirically and established theoretically in \cite{pdmadsssfilterderivative}.
In on-line scenarios where one can admit a random computational complexity
at each time step, the method presented in \cite{olsson2014} is an
interesting alternative when it is applicable.
Empirically, these on-line ML methods converge rather slowly and will
be primarily useful for large data sets.

In a Bayesian framework, batch inference can be conducted using particle
MCMC methods \cite{andrieu2010,Lindsten2014}. However, these
methods are computationally expensive as, for example, an efficient
implementation of the PMMH has a computational complexity of order
$\mathcal{O}(T^{2})$ per iteration \cite{doucet2012}. On-line Bayesian
inference remains a challenging open problem as all methods currently
available, including particle methods with MCMC moves \cite
{Car10,Fea02,Sto02}, suffer from the degeneracy problem.
These methods should not be ruled out, but should be used cautiously,
as they can provide unreliable results even in simple scenarios as
demonstrated in our experiments.

Very recent papers in this dynamic research area have proposed to
combine individual parameter estimation techniques so as to design
more efficient inference algorithms. For example, \cite{Dahlin2014}
suggests to use the score estimation techniques developed for ML parameter
estimation to design better proposal distributions for the PMMH algorithm,
whereas \cite{fearnhead2014} demonstrates that particle methods with
MCMC moves might be fruitfully used in batch scenarios when plugged
into a particle MCMC scheme.




\section*{Acknowledgments}
N. Kantas supported in part by the Engineering and Physical Sciences Research
Council (EPSRC) under Grant EP/J01365X/1 and programme grant on Control
For Energy and Sustainability (EP/G066477/1).
S.~S.~Singh was supported by the EPSRC (grant number EP/G037590/1).
A.~Doucet's research funded in part by EPSRC (EP/K000276/1 and EP/K009850/1).
N.~Chopin's research funded in part by the ANR as part of the ``Investissements
d'Avenir'' program (ANR-11-LABEX-0047).


%

%
\end{document}